\documentclass[amsmath,amssymb,nofootinbib,notitlepage,prd,superscriptaddress,twocolumn]{revtex4-2}
\usepackage[T1]{fontenc}
\usepackage[usenames,dvipsnames]{xcolor}
\usepackage{graphicx,hyperref,orcidlink,float}
\hypersetup{colorlinks=true,citecolor=RoyalBlue,linkcolor=RoyalBlue,urlcolor=RoyalBlue}

\usepackage{appendix}
\usepackage{graphicx}

\usepackage{tikz}
\usepackage{pgfplots}
\usepackage{dcolumn}
\usepackage{bm}
\usepackage{amsmath}

\usepackage[mathlines]{lineno}
\usepackage{xcolor}
\usepackage{slashed}

\usepackage[commentmarkup=uwave,commandnameprefix=ifneeded]{changes}

\usepackage[normalem]{ulem}

\usepackage[normalem]{ulem}

\pgfplotsset{compat=1.18}
\begin{document}

\title{Imprints of Changing Mass and Spin on Black Hole Ringdown}

\newcommand{\Caltech}{\affiliation{Theoretical Astrophysics, Walter Burke
  Institute for Theoretical Physics, California Institute of Technology,
  Pasadena, California 91125, USA}}
\newcommand{\Perimeter}{\affiliation{Perimeter Institute for Theoretical Physics, 31 Caroline Streeth North, Waterloo, Onatrio NSL 2Y5, Canada}}
\newcommand{\PGI}{\affiliation{Princeton Gravity Initiative, Princeton University, Princeton, New Jersey, 08544, USA}}
\newcommand{\Cornell}{\affiliation{Cornell Center for Astrophysics and Planetary Science, Cornell University, Ithaca, New York 14853, USA}}
\newcommand{\AEI}{\affiliation{Max Planck Institute for Gravitational Physics (Albert Einstein Institute), Am M\"uhlenberg 1, D-14476 Potsdam, Germany}}
\newcommand{\Princeton}{\affiliation{Department of Physics, Princeton University, Jadwin Hall, Washington Road, New Jersey, 08544, USA}}
\newcommand{\UIUC}{\affiliation{Illinois Center for Advanced Studies of the Universe \& Department of Physics, University of Illinois at Urbana-Champaign, Urbana, IL 61801, USA}}
\newcommand{\Oberlin}{\affiliation{Department of Physics and Astronomy, Oberlin College}}

\author{Hengrui Zhu \orcidlink{0000-0001-9027-4184}}
\email{hengrui.zhu@princeton.edu}
\Princeton
\PGI

\author{Frans Pretorius}
\Princeton
\PGI


\author{Sizheng Ma \orcidlink{0000-0002-4645-453X}}
\Perimeter


\author{Robert~Owen}
\Oberlin

\newcommand{\CornellPhysics}{\affiliation{Department of Physics,
Cornell University, Ithaca,
    NY, 14853, USA}}

  \newcommand{\MaxPlanck}{\affiliation{Max Planck Institute for
Gravitational
      Physics (Albert Einstein Institute), Am M{\"u}hlenberg 1, D-14476
Potsdam,
      Germany}}

  \author{Yitian Chen \orcidlink{0000-0002-8664-9702}} \Cornell
  \author{Nils Deppe \orcidlink{0000-0003-4557-4115}} \CornellPhysics
\Cornell
  \author{Lawrence E.~Kidder \orcidlink{0000-0001-5392-7342}} \Cornell
  \author{Maria Okounkova} \affiliation{Department of Physics, Pasadena City College, Pasadena, California 91106, USA}
  \author{Harald P.~Pfeiffer \orcidlink{0000-0001-9288-519X}} \MaxPlanck
  \author{Mark A.~Scheel \orcidlink{0000-0001-6656-9134}} \Caltech
  \author{Leo C. Stein \orcidlink{0000-0001-7559-9597}} 
  \affiliation{Department of Physics and Astronomy, University of Mississippi, University, MS 38677, USA}

\date{\today}

\begin{abstract}
We numerically investigate the imprints of gravitational radiation-reaction driven changes to a black hole's mass and spin on the corresponding ringdown waveform.
We do so by comparing the dynamics of a perturbed black hole evolved with the full (nonlinear) versus linearized Einstein equations. As expected, we find that the quasinormal mode amplitudes extracted from nonlinear evolution deviate from their linear counterparts at third order in initial perturbation amplitude. 
For perturbations leading to a change in the black hole mass and spin of $\sim 5\%$, which is reasonable for a remnant formed in an astrophysical merger, we find that nonlinear distortions to the complex amplitudes of some quasinormal modes can be as large as $\sim 50\%$ at the peak of the waveform. 
Furthermore, the change in the mass and spin results in a drift in the quasinormal mode frequencies, which for large amplitude perturbations causes the nonlinear waveform to rapidly dephase with respect to its linear counterpart.
Surprisingly, despite these nonlinear effects creating significant deviations in the nonlinear waveform, we show that a linear quasinormal mode model still performs quite well from close to the peak amplitude onwards.  
\end{abstract}

\maketitle

\section{Introduction}

According to the theory of general relativity, binary black holes undergo an inspiral and merging process, generating ripples in spacetime known as gravitational waves.
These waves carry valuable information about strong-field gravitational dynamics, and are now routinely detected by the LIGO-Virgo-KAGRA collaboration~\cite{AdvancedLIGOScientific:2014pky,VIRGO:2014yos,KAGRA:2020tym}.
Subsequent to the coalescence of the two distinct black hole horizons into one, a phase known as the ringdown occurs, signifying the conclusion of the merger.
In this phase, the remnant black hole settles down to an equilibrium state, emitting gravitational waves with distinctive frequencies that provide insights into the black hole's properties~\cite{Vishveshwara:1970zz,Press:1971wr,Berti:2009kk}.

At some sufficiently late time in the merger process, the ringdown waveform ought to be well described by linear black hole perturbation theory
~\cite{Regge:1957td,Zerilli:1970se,Teukolsky:1973ha}. 
In particular, the Teukolsky equation, governing dynamics of linear perturbations of Kerr black holes, admit eigen solutions known as quasinormal modes (QNMs), which manifest as damped sinusoids with frequencies uniquely determined by the mass and spin of the background Kerr geometry. 
One can then in principle independently measure the remnant mass and spin by fitting the ringdown signal with these QNMs~\cite{Detweiler:1980gk,Dreyer:2003bv,Berti:2005ys}. 
However, there has been much debate regarding how soon after merger one can start the fit using a linear model of ringdown, and how many overtones can faithfully be measured (whether from simulations or actual data)~\cite{Buonanno:2006ui,Berti:2007fi,Giesler:2019uxc,Carullo_GW150914, 
Cotesta:2022pci,Crisostomi:2023tle,Gennari:2023gmx,Correia:2023bfn,PhysRevLett.131.169001,PhysRevLett.131.169002,Finch:2022ynt,Wang:2023xsy,Ma:2023vvr,Ma:2023cwe,Wang:2023mst,Isi:2022mhy,CalderonBustillo:2020rmh,
GW190521_Properties, Capano_GW190521,Nee:2023osy,Qiu:2023lwo}.

The debate is spurred by several complications that arise in the analysis. First, QNMs are not a complete description of black hole perturbations even at the linear level (see e.g. the discussions in~\cite{Ansorg:2016ztf,Green:2022htq}). 
In contrast with for example certain quantum mechanics problems, where the self-adjoint Hamiltonian operator yields a set of eigenmodes that form a complete and orthonormal basis, the QNM solutions of a perturbed black hole are not complete, due in part to the dissipative nature of the theory.~\footnote{Though some conjectures for completeness are formulated with specific boundary conditions, see, e.g. Ref~\cite{London:2023aeo,London:2023idh}.}
Moreover, the transient, typically representing the high frequency non-QNM content of the perturbation, and the power-law tail, arising from back-scattering of the gravitational radiation off spacetime curvature, are present in the ringdown waveform for generic initial data~\cite{Leaver:1985ax,Nollert:1999ji,berti:2006kk,Albanesi:2023bgi,Carullo:2023tff}. Their presence may bias the fitting  using a model that only takes the QNMs into account~\cite{Baibhav:2023clw,Zhu:2023mzv,Takahashi:2023tkb}. 

Furthermore, a plethora of nonlinear effects may dominate the ringdown waveform at early time. 
At second order, linear quasinormal modes couple to give rise to quadratic quasinormal modes~\cite{Gleiser:1995gx,Brizuela:2009qd,Ripley:2020xby,Ioka:2007ak,Nakano:2007cj,Pazos:2010xf,London:2014cma,Lagos:2022otp,Khera:2023lnc,Redondo-Yuste:2023seq,Zhu:2024rej,Ma:2024qcv,Zlochower:2003yh}, which can be louder than their linear counterparts in certain angular harmonics following binary mergers \cite{Mitman:2022qdl,Cheung:2022rbm,cheung2023extracting,Ma:2022wpv,Yi:2024elj}.

Another nonlinear effect that has gained recent attention is the change in the black hole mass and spin due to radiation back-reaction~\cite{Sberna:2021eui,Cannizzaro:2023jle,Redondo-Yuste:2023ipg}. Given that in comparable mass mergers $\sim 5\%$ of the energy of the binary is radiated in gravitational waves in the coalescence and early ringdown phase~\cite{Pretorius:2005gq},
one could expect a significant fraction of that would register as a change in the remnant mass and spin over the corresponding time scale of a few light crossing times. After that, the remnant black hole's properties stabilize exponentially. 

At the perturbative level, changes in the mass and spin can be introduced at the linear level through corresponding zero frequency modes. However, since these modes
are zero frequency, they do not represent changes to the background geometry due to dynamical perturbations (including QNMs and transients). The latter only cause back reaction at second order in perturbation theory, and hence will only produce an effective change in properties of linear QNMs at third order. 

Since even second order perturbations of Kerr black holes are not yet well understood (see e.g.~\cite{Green:2019nam,Andersson:2021eqc,Loutrel:2020wbw,Toomani:2021jlo,Spiers:2023cip}),  
works attempting to address imprints of mass and spin change on QNMs have been heuristic in nature. 
Ref~\cite{Sberna:2021eui} explored what they call Absorption-Induced Mode Excitation (AIME) for black holes in asymptotically Anti-de Sitter spacetime, where they showed that the overtones (fast decaying QNMs) can lead to excitation of the fundamental mode (slowest decaying QNM) due to nonlinear back-reaction on the background spacetime; they further calculated the change in fundamental mode amplitude in asymptotically flat Schwarzschild due to AIME using the "sudden mass-change approximation."
Ref~\cite{Cannizzaro:2023jle} developed a theoretical framework to model the back-reaction of a scalar field on the background spacetime, which yields excellent agreement with numerical simulations.
Ref~\cite{Redondo-Yuste:2023ipg} found evidence for time-dependent frequencies in Vaidya spacetimes, which describe accretion of null dust onto a black hole, yielding a prescribed mass function. 
Concurrent to our work, Ref~\cite{May_in_prep} is investigating AIME for asymptotically flat black holes using a perturbative approach. They find that the prograde QNM can mix into the retrograde QNM, potentially yielding observables of direct astrophysical relevance. 

In this letter, we work without the aforementioned assumptions or approximations and directly investigate the imprints of nonlinear absorption during black hole ringdown through numerical experiments. 
To distinguish linear and nonlinear contributions, we adopt a close-limit type framework~\cite{Price:1994pm,Gleiser:1996yc,Andrade:1996pc,Khanna:1999mh,Allen:1998rg,Baker:1999sj}, in which we evolve a single perturbed black hole by numerically integrating both the linearized and nonlinear Einstein equations. 
With a consistent gauge choice and initial data to linear order, we attribute differences in the extracted gravitational waveform to nonlinear effects. 

A similar numerical approach has been adapted in Ref~\cite{Allen:1998rg,Baker:1999sj} for axially symmetric perturbations of a Schwarzschild black hole; here we focus on non-axisymmetric perturbations of a Kerr black hole with dimensionless spin of 0.7, the expected spin for the remnant formed from the quasi-circular inspiral of two non-spinning equal mass black holes~\cite{Pretorius:2005gq,Tichy:2008du}.
Furthermore, in the earlier works the nonlinear waveforms were normalized by the initial ADM mass (i.e. including all the radiation in the initial data), which results in a mismatch of the time axis compared to the linear waveforms. 
Here, we normalize the time axis by the initial mass of the background black hole to give a consistent comparison between linear and non-linear evolution. 
We also present a detailed ringdown analysis in terms of the wave's quasinormal mode content, quantify different nonlinear effects, and discuss their relevance to observations.  

We focus on the dominant quadrupolar ${\ell=m=2}$ harmonics of the waveform, which are traditionally thought to be dominated by linear contributions. Here we show that nonlinear effects can lead to relative changes of order unity in the amplitude and phase of certain QNMs for the largest, astrophysically reasonable perturbations we consider. 
Despite these large nonlinear distortions, we find that the resulting waveform can still be well-modeled by a linear superposition of QNMs. At a first glance this seems counter-intuitive, however ringdown modeling does not make use of theoretical estimates of the amplitude and phase of excited QNMs, and by the time the waveform is in a regime where the fitting algorithm is stable, the instantaneous frequencies of the QNMs have largely settled down to those of the final remnant black hole. Moreover, the time window over which modes can stably be extracted is essentially the same for the linear vs. nonlinear results, suggesting that transients are the main culprit preventing straightforward interpretation of the coalescence/early-ringdown phase of a merger in terms of black hole perturbation theory, linear or otherwise. 

\emph{QNM Conventions.--} In this letter, individual QNMs are denoted by four indices $(p,\ell,m,n)$, following conventions in Ref.~\cite{Isi:2021iql}. QNM angular structure is described by spheroidal harmonics with indices $\ell$ and $m$, while their radial structure is indicated by the index $n$: $n=0$ QNMs are fundamental modes, while $n>0$ are overtones, ranked by increasing decay rate. For a given $(\ell,m,n)$, there are two QNMs labeled by $p$, signifying modes with wavefronts co-rotating (prograde, ${p=+}$) or counter-rotating (retrograde, ${p=-}$) with the black hole. We define $A_{(p,\ell,m,n)}$ as the complex amplitude for each QNM, and $\omega_{(p,\ell,m,n)}(M,\chi)$ its complex frequency, which is a function of the black hole mass $M$ and dimensionless spin $\chi$. To distinguish amplitudes of QNMs from linear and nonlinear waveform, we use superscripts, e.g. $A_{(p,\ell,m,n)}^{\rm linear}$ and $A_{(p,\ell,m,n)}^{\rm nonlinear}$. 

\begin{figure*}[ht]%
    \includegraphics[width=2\columnwidth]{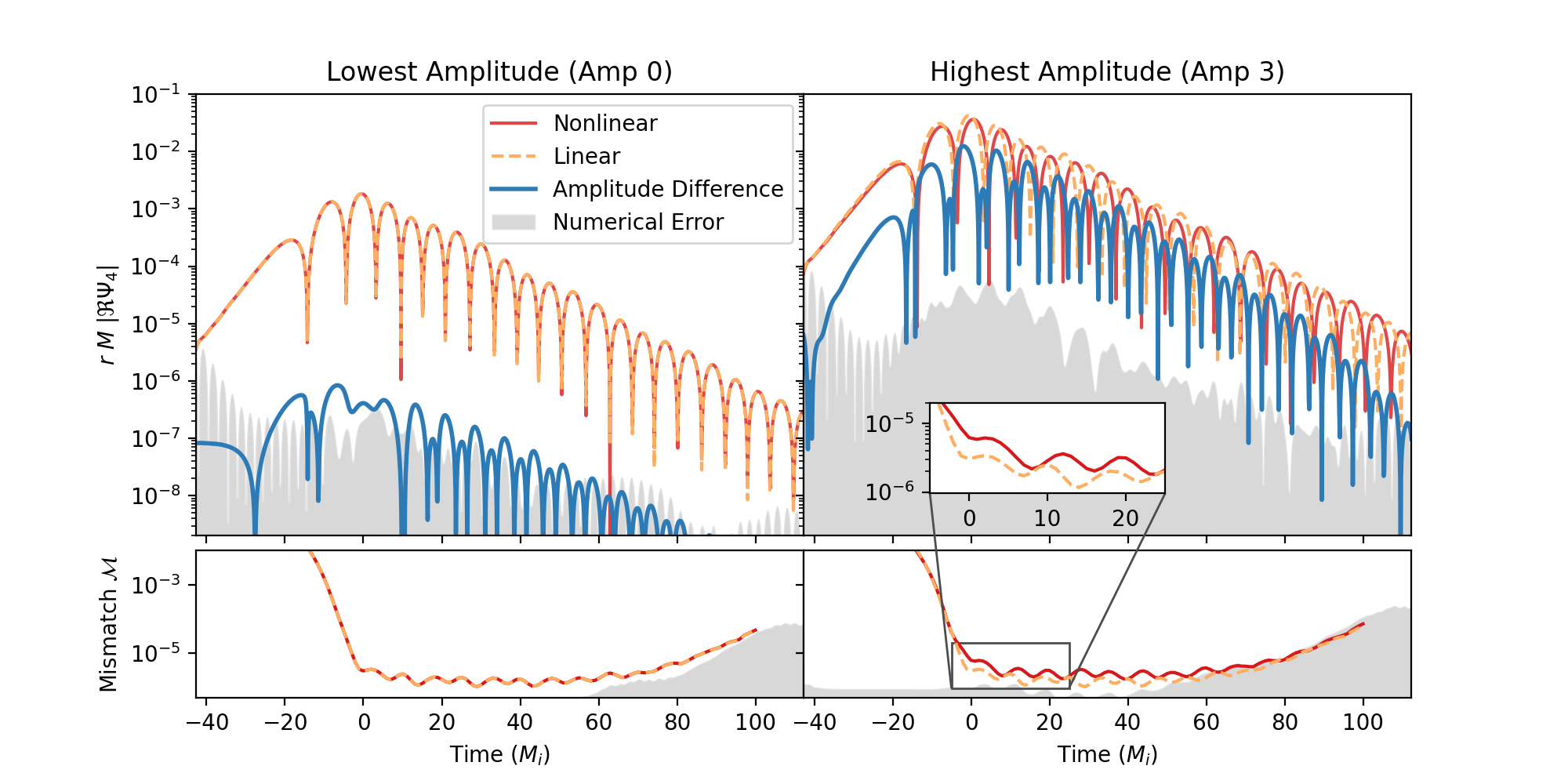}
    \centering
    \caption{
    Top: Comparison between waveforms from linear and nonlinear evolution of a single perturbed black hole. The time axis is normalized by the initial black hole mass $M_i$, and $t=0$ at the peak of the linear waveform. Waveforms from nonlinear evolution are shown as red solid lines and those from linear evolution as orange dashed lines. We further plot the difference between the waveform amplitudes, i.e. $rM\left\vert |\psi_4^{\rm linear}|-|\psi_4^{\rm nonlinear}| \right\vert$, using thick blue lines. We note that the amplitude difference does not take into account the accumulation of a phase difference, which we show in Fig.~\ref{fig:3}. We find good agreement between the nonlinear and linear waveforms at the lowest amplitude (left); however, near order unity relative disagreement (more clearly evident in Fig.~\ref{fig:2}) is found at the highest amplitude (right). The latter causes a change of 3.4\% (5.6\%) in the mass (dimensionless spin) of the black hole in the nonlinear evolution. Bottom: Mismatch between the best fit QNM model (including up to the second overtones) and the waveform varying fitting start time, where we used the late-time mass and spin extracted at the horizon for the nonlinear waveforms. For the highest amplitude case, the mismatch for the nonlinear evolution is only marginally larger than its linear counterpart at late time, suggesting that a linear superposition of QNMs could still be a good model. In all panels, numerical error is estimated using the difference between the highest and second highest resolution nonlinear evolution waveforms; linear evolution shows similar error characteristics.
    }
    \label{fig:1}
\end{figure*}

\section{Numerical Setup}\label{sec:num}
To obtain initial data for a single perturbed black hole, as in Ref~\cite{Zhu:2024rej}, hereby referred to as Paper I, in which we investigated the spin and initial data dependence of quadratic quasinormal modes, we use the conformal thin sandwich approach developed in~\cite{Pfeiffer:2004qz}. First, we superpose the background Kerr metric in spherical Kerr-Schild coordinates $g^0_{ij}$~\cite{Chen:2021rtb} with the following initial guess $h^0_{ij}$ for the metric perturbation: 
\begin{align}
    h^0_{ij}(t,r,\theta,\phi)|_{t=0} &= A\exp\left(-\frac{(r-r_0)^2}{w^2}\right) Y^{lm}_{ij}(\theta,\phi)~,~\\
    \frac{\partial{h}^0_{ij}(t,r,\theta,\phi)}{\partial{t}}|_{t=0} &= 0~,
\end{align}
where $A$, $r_0$, and $w$ are the amplitude, radial location, and width of the initial perturbation respectively, and $Y^{lm}_{ij}$ is the pure-spin tensor spherical harmonics following conventions in, e.g. Ref~\cite{Martel:2005ir}. Here (in contrast to Paper I) we are choosing time-symmetric initial data~\footnote{A time-symmetric pulse can be thought of as a superposition of the ingoing and outgoing pulse as used in Paper I. The outgoing pulse is cleanly separated from the ingoing one by $\delta t = 2r_0$ in the gravitational waveform (see Fig.~\ref{fig:s3} in the appendix), and we ignore it in the analysis here. }. This trivially solves the momentum constraints, but more importantly for our purposes means the background shift vector is not modified, making for more straight-forward comparison between the linear and non-linear evolution.
The Hamiltonian constraint is then solved for a conformal factor $\psi$ rescaling the initial guess for the spatial metric, after which the perturbation can be read off as $h_{ij} = \psi^4(g^0_{ij}+h^0_{ij})-g^0_{ij}$. This procedure yields consistent initial data for the nonlinear and linearized evolutions. 

As mentioned in the introduction, we restrict the initial guess for the perturbation to only have ${\ell = |m| = 2}$ harmonics in the angular sector, and the background black hole to have a dimensionless spin of 0.7 initially. We further fix $r_0= 30 M$ and $w=2M$, though we have checked that our results are largely insensitive to such choices. 

We then numerically evolve this family of initial data, varying $A$ from the perturbative to the non-perturbative regime, using the Spectral Einstein Code (\texttt{SpEC}) for both the linear and nonlinear evolution. 
The nonlinear evolution employs the generalized harmonic formulation of the Einstein equations~\cite{Lindblom:2005qh,Scheel:2006gg,Pretorius:2004jg}. 
For the linearized evolution, we use the implementation by Ref~\cite{Okounkova:2018pql,Okounkova:2019dfo}, which is a consistent linearization of the generalized harmonic equations about the given black hole background. 
In both cases we use a freezing gauge condition, where the gauge source functions are set to be those of the Kerr-Schild background. The first-order gauge source function for the linearized evolution is set to zero for consistency.

\section{Results}\label{sec:res}

\begin{figure}[t]
    \includegraphics[width=\columnwidth]{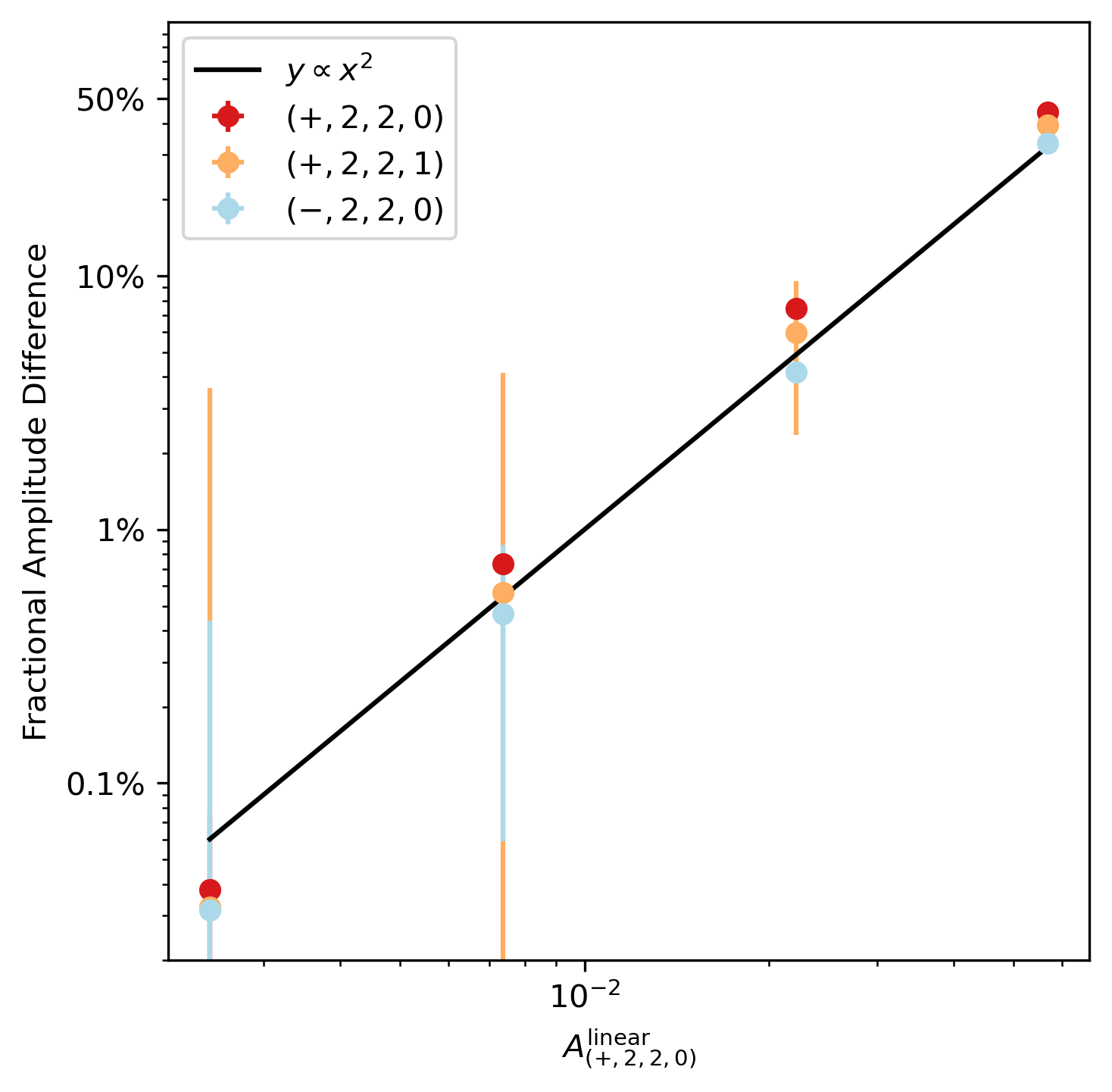}
    \centering
    \caption{
    Shift of the complex QNM amplitudes due to nonlinear effects. On the vertical axis we plot the fractional amplitude difference between the linear and nonlinear waveform for the dominant QNMs $\vert A^{\rm linear}_{(p,l,n,m)}-A^{\rm nonlinear}_{(p,l,n,m)}\vert/\vert A^{\rm linear}_{(p,l,n,m)}\vert$. On the horizontal axis we used the linear $A^{\rm linear}_{(+,2,2,0)}$ as a proxy for linear perturbation amplitude. We find that the QNM amplitudes drift from the linear approximation at third order in the amplitude (hence the fractional amplitudes drift at second order), in accordance with formal perturbation theory. 
    Despite being a higher order effect, for the largest amplitude perturbation we consider, relevant to astrophysical mergers, the relative amplitude changes are close to $\sim50\%$. We stress that this drift in complex amplitude comes from both the absolute amplitude and the phase. 
    }
    \label{fig:2}
\end{figure}

\begin{figure}[t]
    \includegraphics[width=\columnwidth]{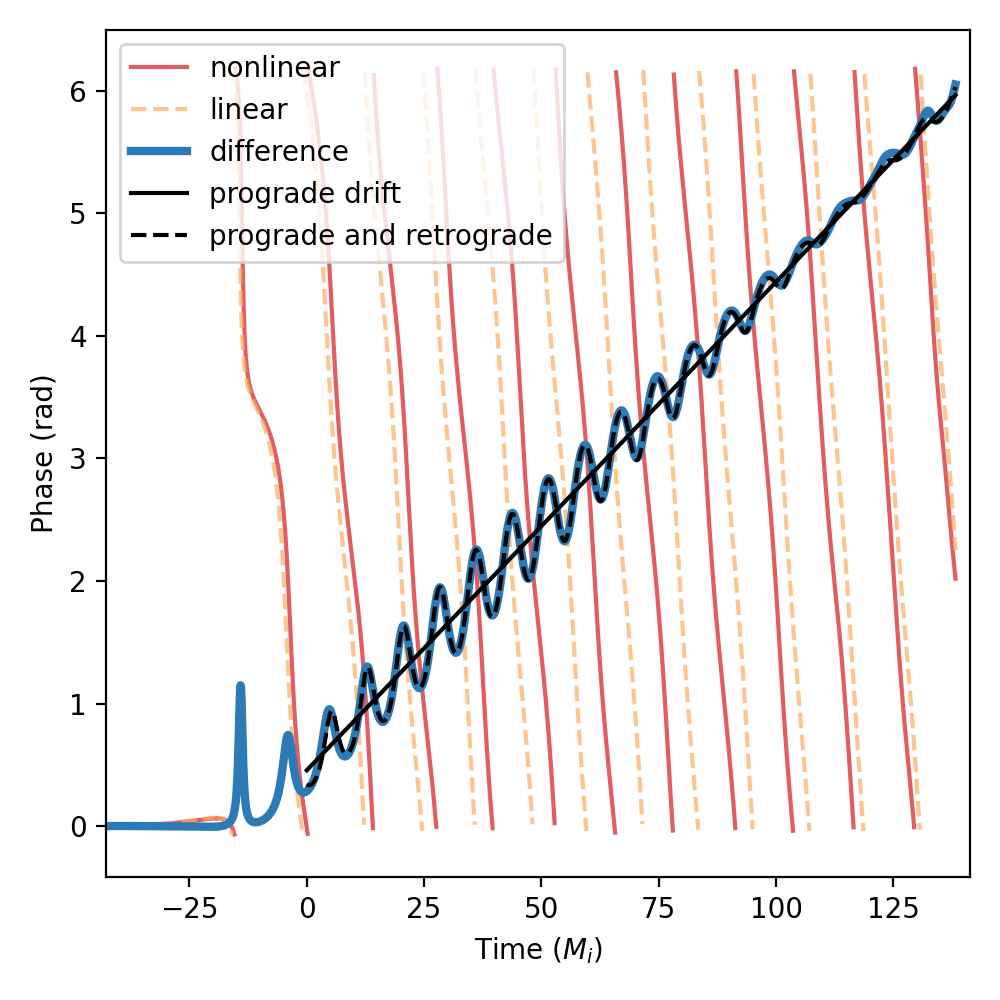}
    \centering
    \caption{Phase difference between the highest amplitude (Amp 3) linear and nonlinear waveform. The phase of the nonlinear waveform is the solid red line, the linear waveform the dashed orange line, and their difference the solid blue line. Following the peak of the waveform ($t=0$), the phase difference can be characterised by a linear drift overlayed with an oscillation that decays away at late time. The linear drift is caused by the change in the $(+,2,2,0)$ frequency due to the mass and spin change; the solid black line is a ``prediction'' of this effect using the corresponding QNM frequencies. The oscillation is caused by the drifted retrograde $(-,2,2,0)$ mode beating against the prograde mode, which decays with time since the retrograde mode decays faster than the prograde mode. We include this effect in the prediction of the difference as the dashed black line. 
    }
    \label{fig:3}
\end{figure}

We performed linear and nonlinear evolutions of perturbed Kerr initial data with 4 amplitudes evenly spaced on a logarithmic scale, labeled with Amp 0-3 from low to high. Amp 0 results in a relative mass and spin change of less than $10^{-4}$, while Amp 3 causes a change in the mass of 3.4\% and dimensionless spin of 5.6\%, where we use the Christodoulou mass and dimensionless spin calculated through approximate Killing fields \cite{Cook:2007wr,Owen:2007dya}. 
In Fig.~\ref{fig:1} we show the amplitude of the linear and nonlinear ${\ell=m=2}$ spin-weighted spherical harmonic component of the waveform extracted at a coordinate radius of 100M with the smallest (Amp 0) and largest (Amp 3) amplitude initial data.~\footnote{We do not use Cauchy-Characteristic Extraction (CCE) as in Paper I since a consistent linearized CCE module does not exist in \textsc{SpECTRE}~\cite{spectrecode}. As shown in Paper I, extracting the wave at a large finite radius with our gauge choice yields consistent results with those calculated through CCE.} 
For the largest Amp 3 run, the ${\ell=m=2}$ waveform receives a distortion in the waveform amplitude of $\sim 30\%$. At a first glance this is surprising in that the amplitude distortion is a third order effect (see Fig.\ref{fig:2}). Though what this indicates is that a perturbation resulting in a $3.4\%$ change in the black hole mass is well into the nonlinear regime, consistent with the expectation that the analogous ``perturbation'' (as measured by mass change) caused by a black hole merger should be considered a rather nonlinear excitation of the remnant.

Before describing the nonlinear changes to the phase evolution of the waveforms, it would be useful to understand their QNM content. Using the methods described in Paper I, we fit the ${\ell=m=2}$ component of the linear and nonlinear waveforms with superpositions of QNMs including up to the second overtone, using the final mass and spin to define the corresponding QNM frequencies. We then compute the mismatch as a function of fitting start time between the best-fit QNM model $\psi_{22}^{\rm QNM}$ and the waveform $\psi_{22}^{\rm NR}$, defined as:
\begin{equation}
    \mathcal{M} = 1-\Re\left\{\frac{\langle \psi^{\rm NR}_{22}\vert \psi^{\rm QNM}_{22}\rangle}{\sqrt{\langle \psi^{\rm NR}_{22}\vert \psi^{\rm NR}_{22}\rangle\langle \psi^{\rm QNM}_{22}\vert \psi^{\rm QNM}_{22}\rangle}}\right\}~,
\end{equation}
where 
\begin{equation}
    \langle a\vert b \rangle = \int_{t=t_0}^{t_f} a(t)\overline{b(t)}{\rm d}t
\end{equation}
In the bottom row of Fig.~\ref{fig:1}, we plot this mismatch as a function of $t_0$ for both the linear and nonlinear waveforms. At early time the mismatch for both the linear and nonlinear waveform is high, which is as expected as here the signal is dominated by a transient. From a time close to the peak amplitude in the waveform onward, the mismatch drops significantly. Surprisingly, the improvement in mismatch, both in magnitude and when it occurs, is largely insensitive to nonlinearities. This suggests that the time when nonlinearities are most relevant in modifying the properties of the QNM excitation is also when the transient is still a significant component of the signal, and hence it's the latter that is ultimately the limiting factor in how early on in a signal a QNM analysis can be used to interpret the ringdown. 

Fig.~\ref{fig:2} illustrates the nonlinear distortion to the complex QNM amplitudes extracted from the linear and nonlinear waveforms for three dominant modes: the fundamental prograde and retrograde modes $(\pm,2,2,0)$, and the first overtone for the prograde mode $(+,2,2,1)$. 
The QNM amplitudes are extrapolated to the peak of the waveform, with uncertainties estimated from stability of fit \cite{Zhu:2024rej,Zhu:2023fnf}. 
What is shown there are the relative differences between nonlinear vs linear evolution as a function of amplitude (see the Supplemental material for plots of the absolute amplitudes, and the stability of the QNM fitting with time). The scaling of the difference with amplitude is as expected from perturbation theory.

Finally, we return to the drift in waveform phase due to nonlinear effects. We illustrate this in Fig.~\ref{fig:3} by showing the phase difference (blue line) between the highest amplitude Amp 3 nonlinear (red line) and linear (dashed orange line) waves versus time.
At early times the two cases are in phase, but then around the time corresponding to the peak amplitude ($t=0$) there is a rapid shift in frequency difference between the two. This results in a phase difference that afterward is dominated by a linear drift in time, the slope of which is consistent with a change of the prograde mode frequency due to the change in mass and spin of the black hole; i.e., the slope of the solid black line is $\Re\left\{\omega_{(+,2,2,0)}(M_i,\chi_i)-\omega_{(+,2,2,0)}(M_f,\chi_f)\right\}$, with $M_i$ and $\chi_i$ ($M_f$ and $\chi_f$) the initial (final) mass and spin. Subdominant to the linear phase drift is a small oscillation that eventually decays away. The oscillation appears to come from beating of the prograde $(+,2,2,0)$ mode with the frequency and amplitude-shifted retrograde mode $(-,2,2,0)$. This oscillation eventually decays away as the retrograde mode decays faster than the prograde mode. The dashed black line in the figure combines this beating effect with the linear drift,
giving an excellent match to the  late-time full nonlinear waveform.

\section{Discussion and Conclusion}\label{sec:conclude}
We investigated nonlinear effects in the gravitational wave perturbation of a black hole by comparing evolution of the same initial perturbation solving the full Einstein equations versus the Einstein equations linearized about the given black hole background. Our main goal for doing this was to investigate how perturbations that results in astrophysically relevant changes in the mass and spin of the black hole affect the QNM excitation, and their extraction from the full signal using models based on linear QNMs. For such large amplitude perturbations we find an order unity relative change in the amplitude of the dominant quadrupolar QNM excited. Though despite such a large difference, we find almost no adverse affect in extracting this mode, its retrograde counterpart and its first overtone, using a fitting procedure based on the linear problem. In otherwords, the time in the fitting procedure where the linear waveform becomes stable is almost identical to the time when it becomes stable for the nonlinear waveform, and the mismatch in either waveform compared to the relevant sum of linear QNMs extracted becomes similarly small. The reason for this appears to be that the time when the nonlinear interaction is most relevant in affecting the amplitude of the QNMs, and indirectly their frequencies through the change in the black hole's mass and spin, is also when the transient part of the signal obscures the interpretation of the waveform as a sum of QNMs. 

Though the particular form of the transient is a function of our initial data, it is difficult to imagine that an effective transient
in the analogous problem of a remnant formed via a black hole merger will be any less significant. The ``good news'' in that regard is our work implies that linear QNM fitting for the ringdown in black hole mergers, whether from numerical predictions or detected events, does not need to model a changing mass and spin. The ``bad news'' is that even were a model of the changing mass and spin developed, it might not help in extending the time over which a stable extraction of QNMs can be obtained due to obscuration by the transient, which is already present at the linear level.\footnote{Applying frequency agnostic fits to the linear and nonlinear waveform may further ascertain the relevance of such models~\cite{Baibhav:2023clw,Cheung:2022rbm,cheung2023extracting}. } Note that our results certainly do not imply that nonlinear effects cannot be observed in the ringdown. For example, quadratic QNMs are present, though they could similarly be incorporated into a linear QNM model by including the appropriate frequency-doubled modes. A significant caveat on the above broader reaching implications of these results is that we have only studied an initial $\chi=0.7$ spin black hole, and a limited family of initial conditions. 

Our framework can be easily extended to investigate other effects during ringdown, such as the quadratic QNMs just mentioned, or precession. Regarding the latter, one could simply rotate the pulse relative to the black hole to study precession of the remnant spin due to radiation back-reaction, and test the toy model proposed in Ref~\cite{Zhu:2023fnf}. We leave a study of these extensions, and a more thorough exploration of parameter space to future endeavors.

\section{Acknowledgements}\label{sec:acknowledgements}

We thank Emanuele Berti, Alejandro Cárdenas-Avendaño, William East, Will Farr, Maximiliano Isi, Luis Lehner, Lionel London, Taillte May, Keefe Mitman, Justin Ripley, Harrison Siegel, Nils Siemonsen, and Huan Yang, for helpful discussions regarding various aspects of this project. 
H.Z. especially thank Alejandro Cárdenas-Avendaño and Justin Ripley for discussions at early phase of this project; Will Farr, Maximiliano Isi, and Harrison Siegel for hosting stimulating discussions at the Flatiron institute; and Nils Siemonsen for suggesting time-symmetric initial data. 
The authors are pleased to acknowledge that the work reported on in this paper was substantially performed using the Princeton Research Computing resources at Princeton University which is a consortium of groups led by the Princeton Institute for Computational Science and Engineering (PICSciE) and Office of Information Technology's Research Computing.
This work was supported in part by the Sherman Fairchild Foundation and NSF Grants, PHY-2011968, PHY-2011961, PHY-2309211,  PHY-2309231, OAC-2209656 at Caltech, and NSF Grants No. PHY-2207342 and OAC-2209655 at Cornell.
F.P. acknowledges support from the NSF through the award PHY-2207286.
LCS was supported by NSF CAREER Award PHY--2047382 and a Sloan Foundation Research Fellowship.

\clearpage

\section*{Supplemental Materials}

\begin{figure}[t]
    \includegraphics[width=\columnwidth]{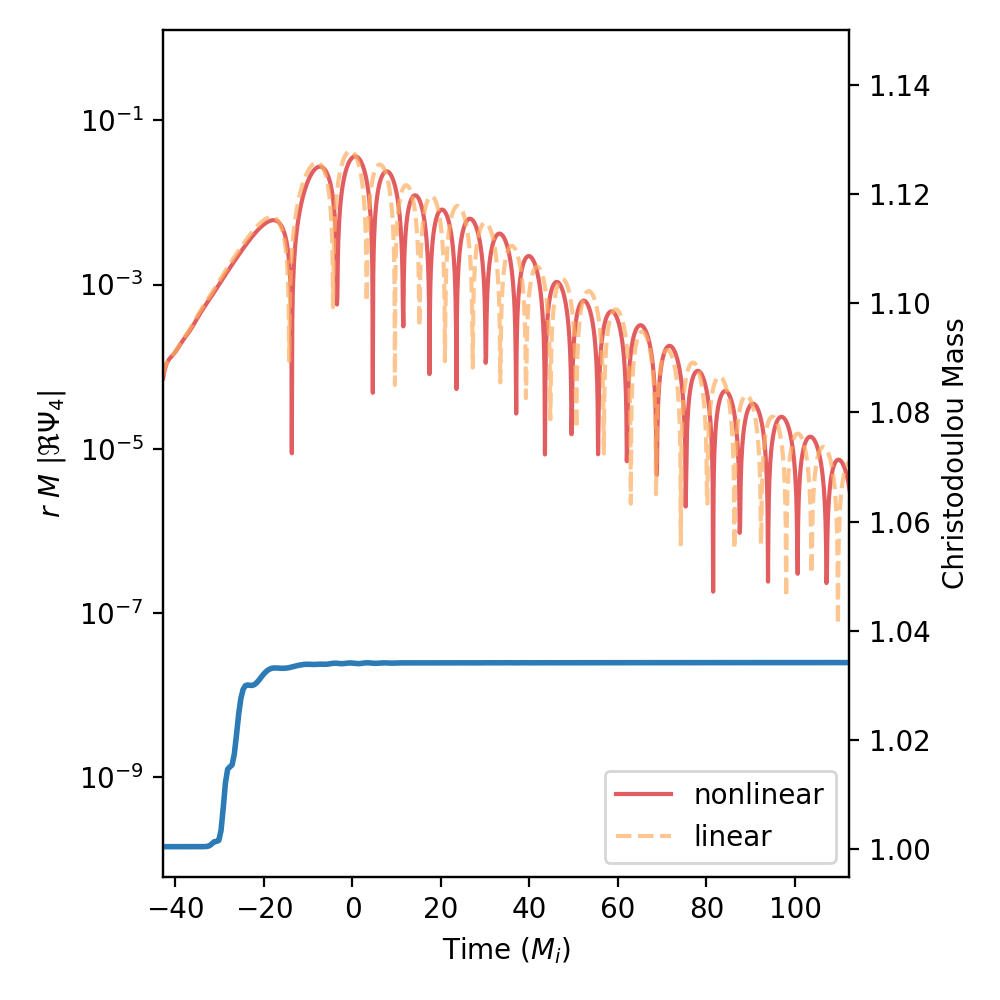}
    \centering
    \caption{An overlay of the waveform and the horizon mass as a function of time for the largest amplitude (Amp 3) run. The horizon time has been shifted by 100 M (the radius of wave-extraction) to align with the waveform. Barring coordinate effects, it appears that the mass change does occur during the transient phase, i.e. before the waveform peaks. Whether this holds true for generic initial data deserves future investigations.  
    }
    \label{fig:s2}
\end{figure}

\begin{figure}[t]
    \includegraphics[width=\columnwidth]{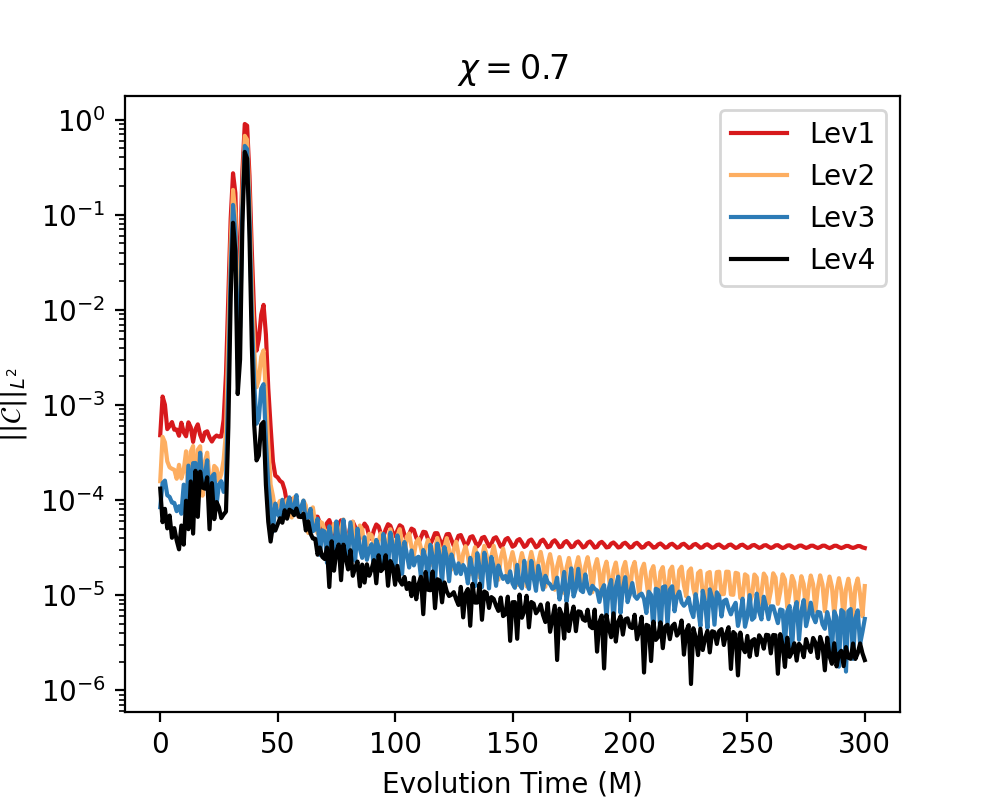}
    \centering
    \caption{$L^2$ norm for the generalized harmonic constraint as a function of simulation time, from the largest amplitude (Amp 3) run. The largest constraint violation is observed when the pulse hits the horizon, which converges away with resolution. 
    }
    \label{fig:s4}
\end{figure}

\begin{figure*}[ht]%
    \includegraphics[width=2\columnwidth]{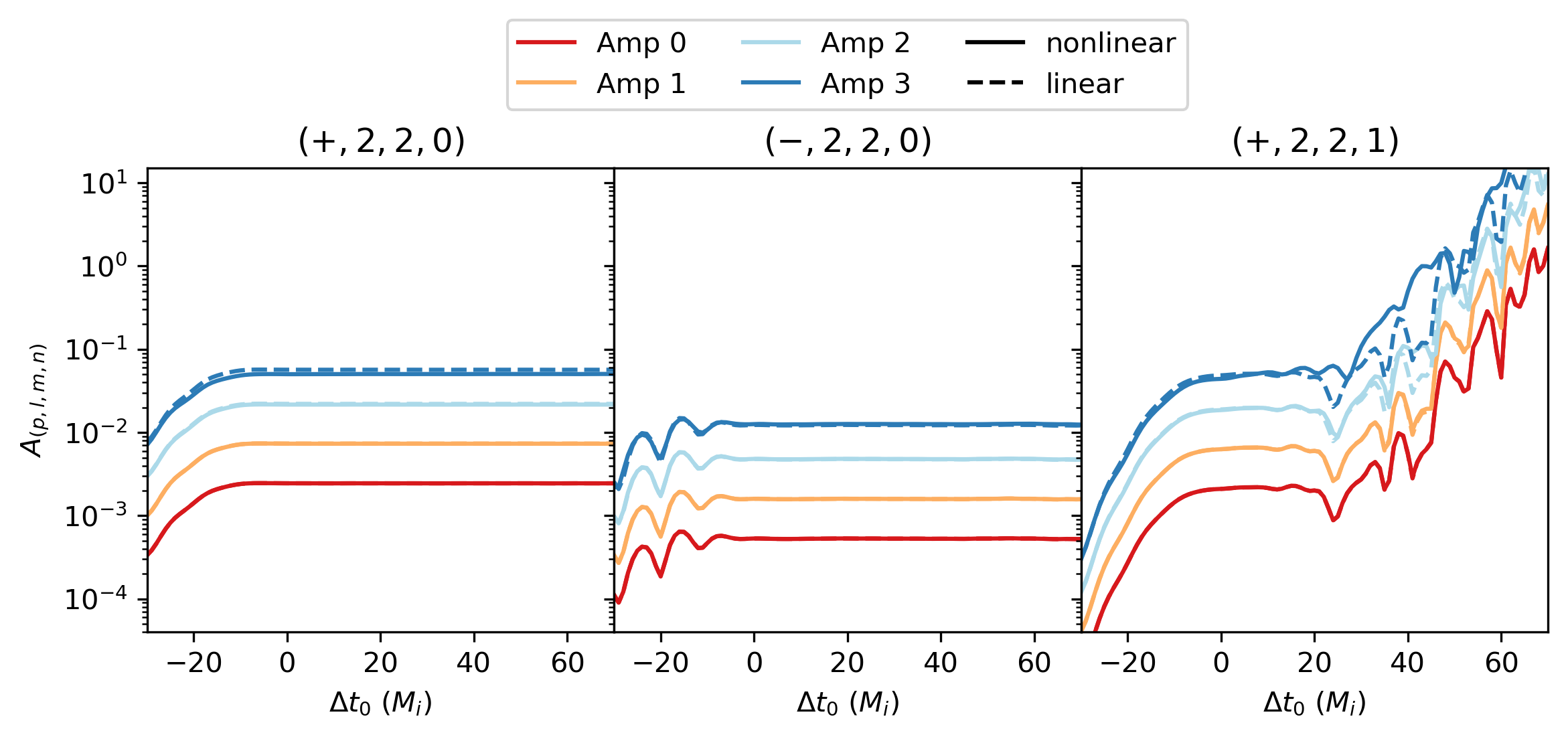}
    \centering
    \caption{Amplitudes of QNMs as a function of fitting start time for the linear (dashed) and nonlinear (solid) waveforms. The QNM is stably extracted if its amplitude remains constant for a sufficiently long period of fit starting time. We note that the nonlinear effects of changing mass and spin does not seem to affect the stability of quasinormal mode fitting, even for the first overtone at the highest amplitude (blue lines in the right most plot). 
    }
    \label{fig:s1}
\end{figure*}

\begin{figure*}[ht]%
    \includegraphics[width=2\columnwidth]{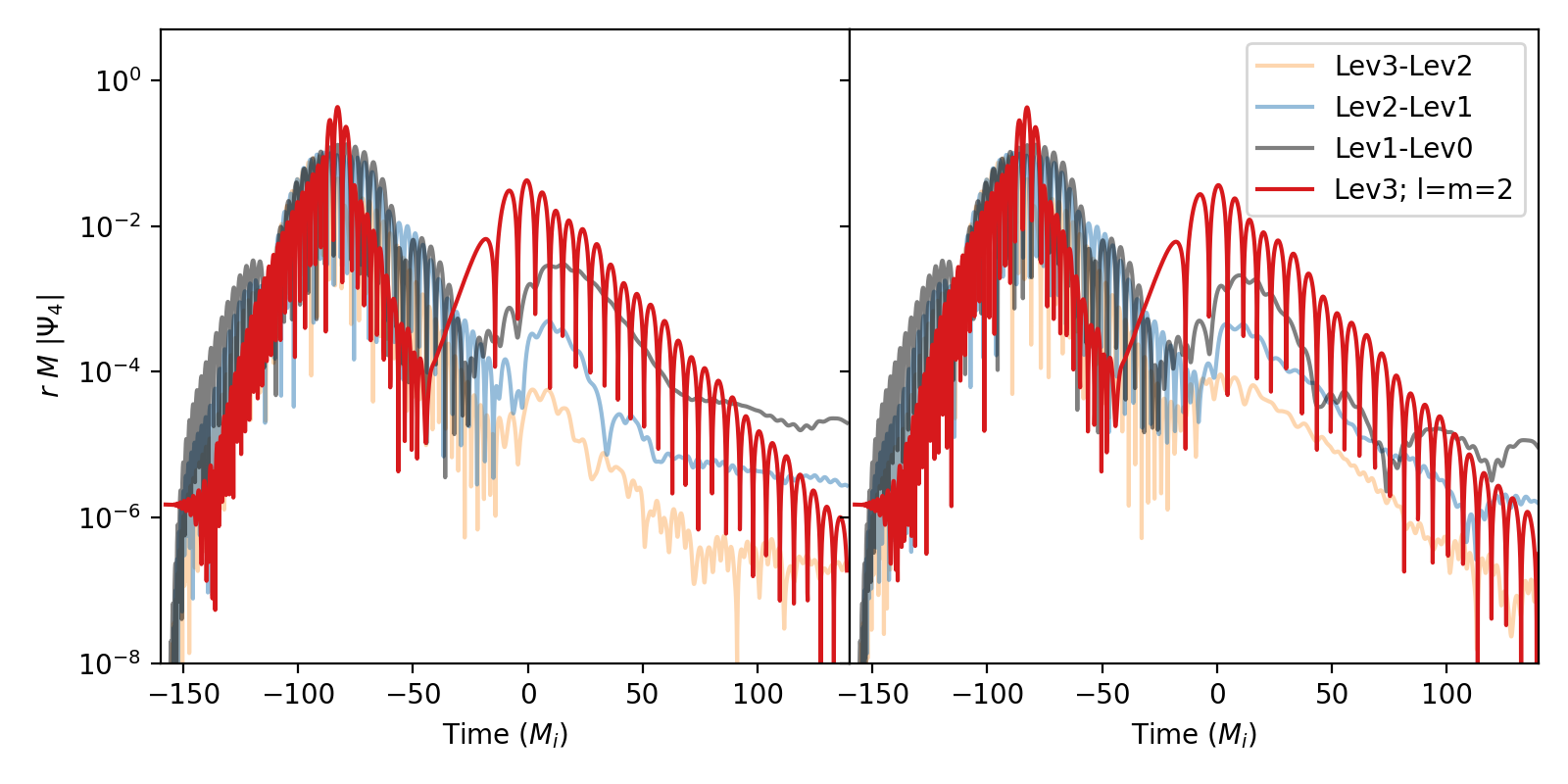}
    \centering
    \caption{
    Convergence of the linear (left) and nonlinear (right) waveform for the Amp3 runs. The resolutions are labeled by Lev0 (lowest) through Lev3 (highest); see Tab.~\ref{tab:s1} for details. We find the residual between different resolutions decreases as expected.
    }
    \label{fig:s3}
\end{figure*}

\appendix
\section{Stability of QNM Fitting}
We refer interested readers to Ref~\cite{Zhu:2023mzv,Zhu:2024rej} for a detailed description of our fitting procedure. 
In Fig.~\ref{fig:s1}, we plot the extracted amplitudes for the three dominant quasinormal modes as a function of fitting start time, for both the linear (dashed line) and nonlinear (solid line) waveforms from all 4 different initial amplitude runs, labeled by color. A QNM is stably extracted if the measured amplitude is nearly constant over a sufficiently large window of start fitting time~\cite{Zhu:2023mzv,Baibhav:2023clw,Takahashi:2023tkb}. We find that even the first overtone (right most panel) is stable for a period longer than 20 M. Surprisingly, the nonlinear effects due to mass and spin change do not seem to affect the stability of the fit even when the change in the background is large. 

In Fig.~\ref{fig:s2}, we plot the horizon mass as a function of time on top of the linear and nonlinear waveform for the highest amplitude (Amp 3) run, where the horizon time is shifted by 100 M to roughly match the waveform time. We note that this matching is very heuristic as the horizon is by definition causally disjoint from the exterior. 
We find that indeed most of the mass change occurs before the peak of the waveform, during which the QNM fitting model fails even for the linear waveform~\cite{Zhu:2023mzv}.

\section{Numerical Convergence}
To test convergence, we run the linear and nonlinear simulations with the highest perturbation amplitude at four different resolutions. The computational domain consists of 24 nested spherical shells with inner boundary at $1 M$ and outer boundary at $330 M$; the number of collocation points in each spherical shell is summarized in Tab.~\ref{tab:s1}. 
We show convergence of the linear and nonlinear Amp 3 waveform in Fig.~\ref{fig:s3}, and convergence of the $L^2$-norm for the generalized harmonic constraint  (see Ref~\cite{Lindblom:2005qh} for its definition) in Fig.~\ref{fig:s4}.

\begin{center}
\begin{table}[h]
\begin{tabular}{|c || c | c|}
 \hline
 Lev & $N_r$ & $N_L$\\
 \hline\hline
 0 & 10 & 14\\ 
 \hline
 1 & 12 & 16\\
 \hline
 2 & 14 & 18\\
 \hline
 3 & 16 & 20\\
 \hline
\end{tabular}
\caption{Number of collocation points/spectral elements for each resolution. }\label{tab:s1}
\end{table}
\end{center}

\clearpage

\bibliography{thebib}

\begin{thebibliography}{93}%
\makeatletter
\providecommand \@ifxundefined [1]{%
 \@ifx{#1\undefined}
}%
\providecommand \@ifnum [1]{%
 \ifnum #1\expandafter \@firstoftwo
 \else \expandafter \@secondoftwo
 \fi
}%
\providecommand \@ifx [1]{%
 \ifx #1\expandafter \@firstoftwo
 \else \expandafter \@secondoftwo
 \fi
}%
\providecommand \natexlab [1]{#1}%
\providecommand \enquote  [1]{``#1''}%
\providecommand \bibnamefont  [1]{#1}%
\providecommand \bibfnamefont [1]{#1}%
\providecommand \citenamefont [1]{#1}%
\providecommand \href@noop [0]{\@secondoftwo}%
\providecommand \href [0]{\begingroup \@sanitize@url \@href}%
\providecommand \@href[1]{\@@startlink{#1}\@@href}%
\providecommand \@@href[1]{\endgroup#1\@@endlink}%
\providecommand \@sanitize@url [0]{\catcode `\\12\catcode `\$12\catcode `\&12\catcode `\#12\catcode `\^12\catcode `\_12\catcode `\%12\relax}%
\providecommand \@@startlink[1]{}%
\providecommand \@@endlink[0]{}%
\providecommand \url  [0]{\begingroup\@sanitize@url \@url }%
\providecommand \@url [1]{\endgroup\@href {#1}{\urlprefix }}%
\providecommand \urlprefix  [0]{URL }%
\providecommand \Eprint [0]{\href }%
\providecommand \doibase [0]{http://dx.doi.org/}%
\providecommand \selectlanguage [0]{\@gobble}%
\providecommand \bibinfo  [0]{\@secondoftwo}%
\providecommand \bibfield  [0]{\@secondoftwo}%
\providecommand \translation [1]{[#1]}%
\providecommand \BibitemOpen [0]{}%
\providecommand \bibitemStop [0]{}%
\providecommand \bibitemNoStop [0]{.\EOS\space}%
\providecommand \EOS [0]{\spacefactor3000\relax}%
\providecommand \BibitemShut  [1]{\csname bibitem#1\endcsname}%
\let\auto@bib@innerbib\@empty
\bibitem [{\citenamefont {Aasi}\ \emph {et~al.}(2015)\citenamefont {Aasi} \emph {et~al.}}]{AdvancedLIGOScientific:2014pky}%
  \BibitemOpen
  \bibfield  {author} {\bibinfo {author} {\bibfnamefont {J.}~\bibnamefont {Aasi}} \emph {et~al.} (\bibinfo {collaboration} {LIGO Scientific}),\ }\href {\doibase 10.1088/0264-9381/32/7/074001} {\bibfield  {journal} {\bibinfo  {journal} {Class. Quant. Grav.}\ }\textbf {\bibinfo {volume} {32}},\ \bibinfo {pages} {074001} (\bibinfo {year} {2015})},\ \Eprint {http://arxiv.org/abs/1411.4547} {arXiv:1411.4547 [gr-qc]} \BibitemShut {NoStop}%
\bibitem [{\citenamefont {Acernese}\ \emph {et~al.}(2015)\citenamefont {Acernese} \emph {et~al.}}]{VIRGO:2014yos}%
  \BibitemOpen
  \bibfield  {author} {\bibinfo {author} {\bibfnamefont {F.}~\bibnamefont {Acernese}} \emph {et~al.} (\bibinfo {collaboration} {VIRGO}),\ }\href {\doibase 10.1088/0264-9381/32/2/024001} {\bibfield  {journal} {\bibinfo  {journal} {Class. Quant. Grav.}\ }\textbf {\bibinfo {volume} {32}},\ \bibinfo {pages} {024001} (\bibinfo {year} {2015})},\ \Eprint {http://arxiv.org/abs/1408.3978} {arXiv:1408.3978 [gr-qc]} \BibitemShut {NoStop}%
\bibitem [{\citenamefont {Akutsu}\ \emph {et~al.}(2021)\citenamefont {Akutsu} \emph {et~al.}}]{KAGRA:2020tym}%
  \BibitemOpen
  \bibfield  {author} {\bibinfo {author} {\bibfnamefont {T.}~\bibnamefont {Akutsu}} \emph {et~al.} (\bibinfo {collaboration} {KAGRA}),\ }\href {\doibase 10.1093/ptep/ptaa125} {\bibfield  {journal} {\bibinfo  {journal} {PTEP}\ }\textbf {\bibinfo {volume} {2021}},\ \bibinfo {pages} {05A101} (\bibinfo {year} {2021})},\ \Eprint {http://arxiv.org/abs/2005.05574} {arXiv:2005.05574 [physics.ins-det]} \BibitemShut {NoStop}%
\bibitem [{\citenamefont {Vishveshwara}(1970)}]{Vishveshwara:1970zz}%
  \BibitemOpen
  \bibfield  {author} {\bibinfo {author} {\bibfnamefont {C.~V.}\ \bibnamefont {Vishveshwara}},\ }\href {\doibase 10.1038/227936a0} {\bibfield  {journal} {\bibinfo  {journal} {Nature}\ }\textbf {\bibinfo {volume} {227}},\ \bibinfo {pages} {936} (\bibinfo {year} {1970})}\BibitemShut {NoStop}%
\bibitem [{\citenamefont {Press}(1971)}]{Press:1971wr}%
  \BibitemOpen
  \bibfield  {author} {\bibinfo {author} {\bibfnamefont {W.~H.}\ \bibnamefont {Press}},\ }\href {\doibase 10.1086/180849} {\bibfield  {journal} {\bibinfo  {journal} {Astrophys. J. Lett.}\ }\textbf {\bibinfo {volume} {170}},\ \bibinfo {pages} {L105} (\bibinfo {year} {1971})}\BibitemShut {NoStop}%
\bibitem [{\citenamefont {Berti}\ \emph {et~al.}(2009)\citenamefont {Berti}, \citenamefont {Cardoso},\ and\ \citenamefont {Starinets}}]{Berti:2009kk}%
  \BibitemOpen
  \bibfield  {author} {\bibinfo {author} {\bibfnamefont {E.}~\bibnamefont {Berti}}, \bibinfo {author} {\bibfnamefont {V.}~\bibnamefont {Cardoso}}, \ and\ \bibinfo {author} {\bibfnamefont {A.~O.}\ \bibnamefont {Starinets}},\ }\href {\doibase 10.1088/0264-9381/26/16/163001} {\bibfield  {journal} {\bibinfo  {journal} {Class. Quant. Grav.}\ }\textbf {\bibinfo {volume} {26}},\ \bibinfo {pages} {163001} (\bibinfo {year} {2009})},\ \Eprint {http://arxiv.org/abs/0905.2975} {arXiv:0905.2975 [gr-qc]} \BibitemShut {NoStop}%
\bibitem [{\citenamefont {Regge}\ and\ \citenamefont {Wheeler}(1957)}]{Regge:1957td}%
  \BibitemOpen
  \bibfield  {author} {\bibinfo {author} {\bibfnamefont {T.}~\bibnamefont {Regge}}\ and\ \bibinfo {author} {\bibfnamefont {J.~A.}\ \bibnamefont {Wheeler}},\ }\href {\doibase 10.1103/PhysRev.108.1063} {\bibfield  {journal} {\bibinfo  {journal} {Phys. Rev.}\ }\textbf {\bibinfo {volume} {108}},\ \bibinfo {pages} {1063} (\bibinfo {year} {1957})}\BibitemShut {NoStop}%
\bibitem [{\citenamefont {Zerilli}(1970)}]{Zerilli:1970se}%
  \BibitemOpen
  \bibfield  {author} {\bibinfo {author} {\bibfnamefont {F.~J.}\ \bibnamefont {Zerilli}},\ }\href {\doibase 10.1103/PhysRevLett.24.737} {\bibfield  {journal} {\bibinfo  {journal} {Phys. Rev. Lett.}\ }\textbf {\bibinfo {volume} {24}},\ \bibinfo {pages} {737} (\bibinfo {year} {1970})}\BibitemShut {NoStop}%
\bibitem [{\citenamefont {Teukolsky}(1973)}]{Teukolsky:1973ha}%
  \BibitemOpen
  \bibfield  {author} {\bibinfo {author} {\bibfnamefont {S.~A.}\ \bibnamefont {Teukolsky}},\ }\href {\doibase 10.1086/152444} {\bibfield  {journal} {\bibinfo  {journal} {Astrophys. J.}\ }\textbf {\bibinfo {volume} {185}},\ \bibinfo {pages} {635} (\bibinfo {year} {1973})}\BibitemShut {NoStop}%
\bibitem [{\citenamefont {Detweiler}(1980)}]{Detweiler:1980gk}%
  \BibitemOpen
  \bibfield  {author} {\bibinfo {author} {\bibfnamefont {S.~L.}\ \bibnamefont {Detweiler}},\ }\href {\doibase 10.1086/158109} {\bibfield  {journal} {\bibinfo  {journal} {Astrophys. J.}\ }\textbf {\bibinfo {volume} {239}},\ \bibinfo {pages} {292} (\bibinfo {year} {1980})}\BibitemShut {NoStop}%
\bibitem [{\citenamefont {Dreyer}\ \emph {et~al.}(2004)\citenamefont {Dreyer}, \citenamefont {Kelly}, \citenamefont {Krishnan}, \citenamefont {Finn}, \citenamefont {Garrison},\ and\ \citenamefont {Lopez-Aleman}}]{Dreyer:2003bv}%
  \BibitemOpen
  \bibfield  {author} {\bibinfo {author} {\bibfnamefont {O.}~\bibnamefont {Dreyer}}, \bibinfo {author} {\bibfnamefont {B.~J.}\ \bibnamefont {Kelly}}, \bibinfo {author} {\bibfnamefont {B.}~\bibnamefont {Krishnan}}, \bibinfo {author} {\bibfnamefont {L.~S.}\ \bibnamefont {Finn}}, \bibinfo {author} {\bibfnamefont {D.}~\bibnamefont {Garrison}}, \ and\ \bibinfo {author} {\bibfnamefont {R.}~\bibnamefont {Lopez-Aleman}},\ }\href {\doibase 10.1088/0264-9381/21/4/003} {\bibfield  {journal} {\bibinfo  {journal} {Class. Quant. Grav.}\ }\textbf {\bibinfo {volume} {21}},\ \bibinfo {pages} {787} (\bibinfo {year} {2004})},\ \Eprint {http://arxiv.org/abs/gr-qc/0309007} {arXiv:gr-qc/0309007} \BibitemShut {NoStop}%
\bibitem [{\citenamefont {Berti}\ \emph {et~al.}(2006)\citenamefont {Berti}, \citenamefont {Cardoso},\ and\ \citenamefont {Will}}]{Berti:2005ys}%
  \BibitemOpen
  \bibfield  {author} {\bibinfo {author} {\bibfnamefont {E.}~\bibnamefont {Berti}}, \bibinfo {author} {\bibfnamefont {V.}~\bibnamefont {Cardoso}}, \ and\ \bibinfo {author} {\bibfnamefont {C.~M.}\ \bibnamefont {Will}},\ }\href {\doibase 10.1103/PhysRevD.73.064030} {\bibfield  {journal} {\bibinfo  {journal} {Phys. Rev. D}\ }\textbf {\bibinfo {volume} {73}},\ \bibinfo {pages} {064030} (\bibinfo {year} {2006})},\ \Eprint {http://arxiv.org/abs/gr-qc/0512160} {arXiv:gr-qc/0512160} \BibitemShut {NoStop}%
\bibitem [{\citenamefont {Buonanno}\ \emph {et~al.}(2007)\citenamefont {Buonanno}, \citenamefont {Cook},\ and\ \citenamefont {Pretorius}}]{Buonanno:2006ui}%
  \BibitemOpen
  \bibfield  {author} {\bibinfo {author} {\bibfnamefont {A.}~\bibnamefont {Buonanno}}, \bibinfo {author} {\bibfnamefont {G.~B.}\ \bibnamefont {Cook}}, \ and\ \bibinfo {author} {\bibfnamefont {F.}~\bibnamefont {Pretorius}},\ }\href {\doibase 10.1103/PhysRevD.75.124018} {\bibfield  {journal} {\bibinfo  {journal} {Phys. Rev. D}\ }\textbf {\bibinfo {volume} {75}},\ \bibinfo {pages} {124018} (\bibinfo {year} {2007})},\ \Eprint {http://arxiv.org/abs/gr-qc/0610122} {arXiv:gr-qc/0610122} \BibitemShut {NoStop}%
\bibitem [{\citenamefont {Berti}\ \emph {et~al.}(2007)\citenamefont {Berti}, \citenamefont {Cardoso}, \citenamefont {Gonzalez}, \citenamefont {Sperhake}, \citenamefont {Hannam}, \citenamefont {Husa},\ and\ \citenamefont {Bruegmann}}]{Berti:2007fi}%
  \BibitemOpen
  \bibfield  {author} {\bibinfo {author} {\bibfnamefont {E.}~\bibnamefont {Berti}}, \bibinfo {author} {\bibfnamefont {V.}~\bibnamefont {Cardoso}}, \bibinfo {author} {\bibfnamefont {J.~A.}\ \bibnamefont {Gonzalez}}, \bibinfo {author} {\bibfnamefont {U.}~\bibnamefont {Sperhake}}, \bibinfo {author} {\bibfnamefont {M.}~\bibnamefont {Hannam}}, \bibinfo {author} {\bibfnamefont {S.}~\bibnamefont {Husa}}, \ and\ \bibinfo {author} {\bibfnamefont {B.}~\bibnamefont {Bruegmann}},\ }\href {\doibase 10.1103/PhysRevD.76.064034} {\bibfield  {journal} {\bibinfo  {journal} {Phys. Rev. D}\ }\textbf {\bibinfo {volume} {76}},\ \bibinfo {pages} {064034} (\bibinfo {year} {2007})},\ \Eprint {http://arxiv.org/abs/gr-qc/0703053} {arXiv:gr-qc/0703053} \BibitemShut {NoStop}%
\bibitem [{\citenamefont {Giesler}\ \emph {et~al.}(2019)\citenamefont {Giesler}, \citenamefont {Isi}, \citenamefont {Scheel},\ and\ \citenamefont {Teukolsky}}]{Giesler:2019uxc}%
  \BibitemOpen
  \bibfield  {author} {\bibinfo {author} {\bibfnamefont {M.}~\bibnamefont {Giesler}}, \bibinfo {author} {\bibfnamefont {M.}~\bibnamefont {Isi}}, \bibinfo {author} {\bibfnamefont {M.~A.}\ \bibnamefont {Scheel}}, \ and\ \bibinfo {author} {\bibfnamefont {S.}~\bibnamefont {Teukolsky}},\ }\href {\doibase 10.1103/PhysRevX.9.041060} {\bibfield  {journal} {\bibinfo  {journal} {Phys. Rev. X}\ }\textbf {\bibinfo {volume} {9}},\ \bibinfo {pages} {041060} (\bibinfo {year} {2019})},\ \Eprint {http://arxiv.org/abs/1903.08284} {arXiv:1903.08284 [gr-qc]} \BibitemShut {NoStop}%
\bibitem [{\citenamefont {Carullo}\ \emph {et~al.}(2019)\citenamefont {Carullo}, \citenamefont {Del~Pozzo},\ and\ \citenamefont {Veitch}}]{Carullo_GW150914}%
  \BibitemOpen
  \bibfield  {author} {\bibinfo {author} {\bibfnamefont {G.}~\bibnamefont {Carullo}}, \bibinfo {author} {\bibfnamefont {W.}~\bibnamefont {Del~Pozzo}}, \ and\ \bibinfo {author} {\bibfnamefont {J.}~\bibnamefont {Veitch}},\ }\href {\doibase 10.1103/PhysRevD.99.123029} {\bibfield  {journal} {\bibinfo  {journal} {Phys. Rev. D}\ }\textbf {\bibinfo {volume} {99}},\ \bibinfo {pages} {123029} (\bibinfo {year} {2019})},\ \bibinfo {note} {[Erratum: Phys.Rev.D 100, 089903 (2019)]},\ \Eprint {http://arxiv.org/abs/1902.07527} {arXiv:1902.07527 [gr-qc]} \BibitemShut {NoStop}%
\bibitem [{\citenamefont {Cotesta}\ \emph {et~al.}(2022)\citenamefont {Cotesta}, \citenamefont {Carullo}, \citenamefont {Berti},\ and\ \citenamefont {Cardoso}}]{Cotesta:2022pci}%
  \BibitemOpen
  \bibfield  {author} {\bibinfo {author} {\bibfnamefont {R.}~\bibnamefont {Cotesta}}, \bibinfo {author} {\bibfnamefont {G.}~\bibnamefont {Carullo}}, \bibinfo {author} {\bibfnamefont {E.}~\bibnamefont {Berti}}, \ and\ \bibinfo {author} {\bibfnamefont {V.}~\bibnamefont {Cardoso}},\ }\href {\doibase 10.1103/PhysRevLett.129.111102} {\bibfield  {journal} {\bibinfo  {journal} {Phys. Rev. Lett.}\ }\textbf {\bibinfo {volume} {129}},\ \bibinfo {pages} {111102} (\bibinfo {year} {2022})},\ \Eprint {http://arxiv.org/abs/2201.00822} {arXiv:2201.00822 [gr-qc]} \BibitemShut {NoStop}%
\bibitem [{\citenamefont {Crisostomi}\ \emph {et~al.}(2023)\citenamefont {Crisostomi}, \citenamefont {Dey}, \citenamefont {Barausse},\ and\ \citenamefont {Trotta}}]{Crisostomi:2023tle}%
  \BibitemOpen
  \bibfield  {author} {\bibinfo {author} {\bibfnamefont {M.}~\bibnamefont {Crisostomi}}, \bibinfo {author} {\bibfnamefont {K.}~\bibnamefont {Dey}}, \bibinfo {author} {\bibfnamefont {E.}~\bibnamefont {Barausse}}, \ and\ \bibinfo {author} {\bibfnamefont {R.}~\bibnamefont {Trotta}},\ }\href {\doibase 10.1103/PhysRevD.108.044029} {\bibfield  {journal} {\bibinfo  {journal} {Phys. Rev. D}\ }\textbf {\bibinfo {volume} {108}},\ \bibinfo {pages} {044029} (\bibinfo {year} {2023})},\ \Eprint {http://arxiv.org/abs/2305.18528} {arXiv:2305.18528 [gr-qc]} \BibitemShut {NoStop}%
\bibitem [{\citenamefont {Gennari}\ \emph {et~al.}(2023)\citenamefont {Gennari}, \citenamefont {Carullo},\ and\ \citenamefont {Del~Pozzo}}]{Gennari:2023gmx}%
  \BibitemOpen
  \bibfield  {author} {\bibinfo {author} {\bibfnamefont {V.}~\bibnamefont {Gennari}}, \bibinfo {author} {\bibfnamefont {G.}~\bibnamefont {Carullo}}, \ and\ \bibinfo {author} {\bibfnamefont {W.}~\bibnamefont {Del~Pozzo}},\ }\href@noop {} {\bibfield  {journal} {\bibinfo  {journal} {arXiv e-prints}\ } (\bibinfo {year} {2023})},\ \Eprint {http://arxiv.org/abs/2312.12515} {arXiv:2312.12515 [gr-qc]} \BibitemShut {NoStop}%
\bibitem [{\citenamefont {Correia}\ \emph {et~al.}(2023)\citenamefont {Correia}, \citenamefont {Wang},\ and\ \citenamefont {Capano}}]{Correia:2023bfn}%
  \BibitemOpen
  \bibfield  {author} {\bibinfo {author} {\bibfnamefont {A.}~\bibnamefont {Correia}}, \bibinfo {author} {\bibfnamefont {Y.-F.}\ \bibnamefont {Wang}}, \ and\ \bibinfo {author} {\bibfnamefont {C.~D.}\ \bibnamefont {Capano}},\ }\href@noop {} {\bibfield  {journal} {\bibinfo  {journal} {arXiv e-prints}\ } (\bibinfo {year} {2023})},\ \Eprint {http://arxiv.org/abs/2312.14118} {arXiv:2312.14118 [gr-qc]} \BibitemShut {NoStop}%
\bibitem [{\citenamefont {Isi}\ and\ \citenamefont {Farr}(2023)}]{PhysRevLett.131.169001}%
  \BibitemOpen
  \bibfield  {author} {\bibinfo {author} {\bibfnamefont {M.}~\bibnamefont {Isi}}\ and\ \bibinfo {author} {\bibfnamefont {W.~M.}\ \bibnamefont {Farr}},\ }\href {\doibase 10.1103/PhysRevLett.131.169001} {\bibfield  {journal} {\bibinfo  {journal} {Phys. Rev. Lett.}\ }\textbf {\bibinfo {volume} {131}},\ \bibinfo {pages} {169001} (\bibinfo {year} {2023})}\BibitemShut {NoStop}%
\bibitem [{\citenamefont {Carullo}\ \emph {et~al.}(2023)\citenamefont {Carullo}, \citenamefont {Cotesta}, \citenamefont {Berti},\ and\ \citenamefont {Cardoso}}]{PhysRevLett.131.169002}%
  \BibitemOpen
  \bibfield  {author} {\bibinfo {author} {\bibfnamefont {G.}~\bibnamefont {Carullo}}, \bibinfo {author} {\bibfnamefont {R.}~\bibnamefont {Cotesta}}, \bibinfo {author} {\bibfnamefont {E.}~\bibnamefont {Berti}}, \ and\ \bibinfo {author} {\bibfnamefont {V.}~\bibnamefont {Cardoso}},\ }\href {\doibase 10.1103/PhysRevLett.131.169002} {\bibfield  {journal} {\bibinfo  {journal} {Phys. Rev. Lett.}\ }\textbf {\bibinfo {volume} {131}},\ \bibinfo {pages} {169002} (\bibinfo {year} {2023})}\BibitemShut {NoStop}%
\bibitem [{\citenamefont {Finch}\ and\ \citenamefont {Moore}(2022)}]{Finch:2022ynt}%
  \BibitemOpen
  \bibfield  {author} {\bibinfo {author} {\bibfnamefont {E.}~\bibnamefont {Finch}}\ and\ \bibinfo {author} {\bibfnamefont {C.~J.}\ \bibnamefont {Moore}},\ }\href {\doibase 10.1103/PhysRevD.106.043005} {\bibfield  {journal} {\bibinfo  {journal} {Phys. Rev. D}\ }\textbf {\bibinfo {volume} {106}},\ \bibinfo {pages} {043005} (\bibinfo {year} {2022})},\ \Eprint {http://arxiv.org/abs/2205.07809} {arXiv:2205.07809 [gr-qc]} \BibitemShut {NoStop}%
\bibitem [{\citenamefont {Wang}\ \emph {et~al.}(2023)\citenamefont {Wang}, \citenamefont {Capano}, \citenamefont {Abedi}, \citenamefont {Kastha}, \citenamefont {Krishnan}, \citenamefont {Nielsen}, \citenamefont {Nitz},\ and\ \citenamefont {Westerweck}}]{Wang:2023xsy}%
  \BibitemOpen
  \bibfield  {author} {\bibinfo {author} {\bibfnamefont {Y.-F.}\ \bibnamefont {Wang}}, \bibinfo {author} {\bibfnamefont {C.~D.}\ \bibnamefont {Capano}}, \bibinfo {author} {\bibfnamefont {J.}~\bibnamefont {Abedi}}, \bibinfo {author} {\bibfnamefont {S.}~\bibnamefont {Kastha}}, \bibinfo {author} {\bibfnamefont {B.}~\bibnamefont {Krishnan}}, \bibinfo {author} {\bibfnamefont {A.~B.}\ \bibnamefont {Nielsen}}, \bibinfo {author} {\bibfnamefont {A.~H.}\ \bibnamefont {Nitz}}, \ and\ \bibinfo {author} {\bibfnamefont {J.}~\bibnamefont {Westerweck}},\ }\href@noop {} {\bibfield  {journal} {\bibinfo  {journal} {arXiv e-prints}\ } (\bibinfo {year} {2023})},\ \Eprint {http://arxiv.org/abs/2310.19645} {arXiv:2310.19645 [gr-qc]} \BibitemShut {NoStop}%
\bibitem [{\citenamefont {Ma}\ \emph {et~al.}(2023{\natexlab{a}})\citenamefont {Ma}, \citenamefont {Sun},\ and\ \citenamefont {Chen}}]{Ma:2023vvr}%
  \BibitemOpen
  \bibfield  {author} {\bibinfo {author} {\bibfnamefont {S.}~\bibnamefont {Ma}}, \bibinfo {author} {\bibfnamefont {L.}~\bibnamefont {Sun}}, \ and\ \bibinfo {author} {\bibfnamefont {Y.}~\bibnamefont {Chen}},\ }\href {\doibase 10.1103/PhysRevD.107.084010} {\bibfield  {journal} {\bibinfo  {journal} {Phys. Rev. D}\ }\textbf {\bibinfo {volume} {107}},\ \bibinfo {pages} {084010} (\bibinfo {year} {2023}{\natexlab{a}})},\ \Eprint {http://arxiv.org/abs/2301.06639} {arXiv:2301.06639 [gr-qc]} \BibitemShut {NoStop}%
\bibitem [{\citenamefont {Ma}\ \emph {et~al.}(2023{\natexlab{b}})\citenamefont {Ma}, \citenamefont {Sun},\ and\ \citenamefont {Chen}}]{Ma:2023cwe}%
  \BibitemOpen
  \bibfield  {author} {\bibinfo {author} {\bibfnamefont {S.}~\bibnamefont {Ma}}, \bibinfo {author} {\bibfnamefont {L.}~\bibnamefont {Sun}}, \ and\ \bibinfo {author} {\bibfnamefont {Y.}~\bibnamefont {Chen}},\ }\href {\doibase 10.1103/PhysRevLett.130.141401} {\bibfield  {journal} {\bibinfo  {journal} {Phys. Rev. Lett.}\ }\textbf {\bibinfo {volume} {130}},\ \bibinfo {pages} {141401} (\bibinfo {year} {2023}{\natexlab{b}})},\ \Eprint {http://arxiv.org/abs/2301.06705} {arXiv:2301.06705 [gr-qc]} \BibitemShut {NoStop}%
\bibitem [{\citenamefont {Wang}\ and\ \citenamefont {Shao}(2023)}]{Wang:2023mst}%
  \BibitemOpen
  \bibfield  {author} {\bibinfo {author} {\bibfnamefont {H.-T.}\ \bibnamefont {Wang}}\ and\ \bibinfo {author} {\bibfnamefont {L.}~\bibnamefont {Shao}},\ }\href {\doibase 10.1103/PhysRevD.108.123018} {\bibfield  {journal} {\bibinfo  {journal} {Phys. Rev. D}\ }\textbf {\bibinfo {volume} {108}},\ \bibinfo {pages} {123018} (\bibinfo {year} {2023})},\ \Eprint {http://arxiv.org/abs/2311.13300} {arXiv:2311.13300 [gr-qc]} \BibitemShut {NoStop}%
\bibitem [{\citenamefont {Isi}\ and\ \citenamefont {Farr}(2022)}]{Isi:2022mhy}%
  \BibitemOpen
  \bibfield  {author} {\bibinfo {author} {\bibfnamefont {M.}~\bibnamefont {Isi}}\ and\ \bibinfo {author} {\bibfnamefont {W.~M.}\ \bibnamefont {Farr}},\ }\href@noop {} {\bibfield  {journal} {\bibinfo  {journal} {arXiv e-prints}\ } (\bibinfo {year} {2022})},\ \Eprint {http://arxiv.org/abs/2202.02941} {arXiv:2202.02941 [gr-qc]} \BibitemShut {NoStop}%
\bibitem [{\citenamefont {Calder\'on~Bustillo}\ \emph {et~al.}(2021)\citenamefont {Calder\'on~Bustillo}, \citenamefont {Lasky},\ and\ \citenamefont {Thrane}}]{CalderonBustillo:2020rmh}%
  \BibitemOpen
  \bibfield  {author} {\bibinfo {author} {\bibfnamefont {J.}~\bibnamefont {Calder\'on~Bustillo}}, \bibinfo {author} {\bibfnamefont {P.~D.}\ \bibnamefont {Lasky}}, \ and\ \bibinfo {author} {\bibfnamefont {E.}~\bibnamefont {Thrane}},\ }\href {\doibase 10.1103/PhysRevD.103.024041} {\bibfield  {journal} {\bibinfo  {journal} {Phys. Rev. D}\ }\textbf {\bibinfo {volume} {103}},\ \bibinfo {pages} {024041} (\bibinfo {year} {2021})},\ \Eprint {http://arxiv.org/abs/2010.01857} {arXiv:2010.01857 [gr-qc]} \BibitemShut {NoStop}%
\bibitem [{\citenamefont {Abbott}\ \emph {et~al.}(2020)\citenamefont {Abbott} \emph {et~al.}}]{GW190521_Properties}%
  \BibitemOpen
  \bibfield  {author} {\bibinfo {author} {\bibfnamefont {R.}~\bibnamefont {Abbott}} \emph {et~al.} (\bibinfo {collaboration} {LIGO Scientific, Virgo}),\ }\href {\doibase 10.3847/2041-8213/aba493} {\bibfield  {journal} {\bibinfo  {journal} {Astrophys. J. Lett.}\ }\textbf {\bibinfo {volume} {900}},\ \bibinfo {pages} {L13} (\bibinfo {year} {2020})},\ \Eprint {http://arxiv.org/abs/2009.01190} {arXiv:2009.01190 [astro-ph.HE]} \BibitemShut {NoStop}%
\bibitem [{\citenamefont {Capano}\ \emph {et~al.}(2023)\citenamefont {Capano}, \citenamefont {Cabero}, \citenamefont {Westerweck}, \citenamefont {Abedi}, \citenamefont {Kastha}, \citenamefont {Nitz}, \citenamefont {Wang}, \citenamefont {Nielsen},\ and\ \citenamefont {Krishnan}}]{Capano_GW190521}%
  \BibitemOpen
  \bibfield  {author} {\bibinfo {author} {\bibfnamefont {C.~D.}\ \bibnamefont {Capano}}, \bibinfo {author} {\bibfnamefont {M.}~\bibnamefont {Cabero}}, \bibinfo {author} {\bibfnamefont {J.}~\bibnamefont {Westerweck}}, \bibinfo {author} {\bibfnamefont {J.}~\bibnamefont {Abedi}}, \bibinfo {author} {\bibfnamefont {S.}~\bibnamefont {Kastha}}, \bibinfo {author} {\bibfnamefont {A.~H.}\ \bibnamefont {Nitz}}, \bibinfo {author} {\bibfnamefont {Y.-F.}\ \bibnamefont {Wang}}, \bibinfo {author} {\bibfnamefont {A.~B.}\ \bibnamefont {Nielsen}}, \ and\ \bibinfo {author} {\bibfnamefont {B.}~\bibnamefont {Krishnan}},\ }\href {\doibase 10.1103/PhysRevLett.131.221402} {\bibfield  {journal} {\bibinfo  {journal} {Phys. Rev. Lett.}\ }\textbf {\bibinfo {volume} {131}},\ \bibinfo {pages} {221402} (\bibinfo {year} {2023})},\ \Eprint {http://arxiv.org/abs/2105.05238} {arXiv:2105.05238 [gr-qc]} \BibitemShut {NoStop}%
\bibitem [{\citenamefont {Nee}\ \emph {et~al.}(2023)\citenamefont {Nee}, \citenamefont {V\"olkel},\ and\ \citenamefont {Pfeiffer}}]{Nee:2023osy}%
  \BibitemOpen
  \bibfield  {author} {\bibinfo {author} {\bibfnamefont {P.~J.}\ \bibnamefont {Nee}}, \bibinfo {author} {\bibfnamefont {S.~H.}\ \bibnamefont {V\"olkel}}, \ and\ \bibinfo {author} {\bibfnamefont {H.~P.}\ \bibnamefont {Pfeiffer}},\ }\href {\doibase 10.1103/PhysRevD.108.044032} {\bibfield  {journal} {\bibinfo  {journal} {Phys. Rev. D}\ }\textbf {\bibinfo {volume} {108}},\ \bibinfo {pages} {044032} (\bibinfo {year} {2023})},\ \Eprint {http://arxiv.org/abs/2302.06634} {arXiv:2302.06634 [gr-qc]} \BibitemShut {NoStop}%
\bibitem [{\citenamefont {Qiu}\ \emph {et~al.}(2024)\citenamefont {Qiu}, \citenamefont {Forteza},\ and\ \citenamefont {Mourier}}]{Qiu:2023lwo}%
  \BibitemOpen
  \bibfield  {author} {\bibinfo {author} {\bibfnamefont {Y.}~\bibnamefont {Qiu}}, \bibinfo {author} {\bibfnamefont {X.~J.}\ \bibnamefont {Forteza}}, \ and\ \bibinfo {author} {\bibfnamefont {P.}~\bibnamefont {Mourier}},\ }\href {\doibase 10.1103/PhysRevD.109.064075} {\bibfield  {journal} {\bibinfo  {journal} {Phys. Rev. D}\ }\textbf {\bibinfo {volume} {109}},\ \bibinfo {pages} {064075} (\bibinfo {year} {2024})},\ \Eprint {http://arxiv.org/abs/2312.15904} {arXiv:2312.15904 [gr-qc]} \BibitemShut {NoStop}%
\bibitem [{\citenamefont {Ansorg}\ and\ \citenamefont {Panosso~Macedo}(2016)}]{Ansorg:2016ztf}%
  \BibitemOpen
  \bibfield  {author} {\bibinfo {author} {\bibfnamefont {M.}~\bibnamefont {Ansorg}}\ and\ \bibinfo {author} {\bibfnamefont {R.}~\bibnamefont {Panosso~Macedo}},\ }\href {\doibase 10.1103/PhysRevD.93.124016} {\bibfield  {journal} {\bibinfo  {journal} {Phys. Rev. D}\ }\textbf {\bibinfo {volume} {93}},\ \bibinfo {pages} {124016} (\bibinfo {year} {2016})},\ \Eprint {http://arxiv.org/abs/1604.02261} {arXiv:1604.02261 [gr-qc]} \BibitemShut {NoStop}%
\bibitem [{\citenamefont {Green}\ \emph {et~al.}(2023)\citenamefont {Green}, \citenamefont {Hollands}, \citenamefont {Sberna}, \citenamefont {Toomani},\ and\ \citenamefont {Zimmerman}}]{Green:2022htq}%
  \BibitemOpen
  \bibfield  {author} {\bibinfo {author} {\bibfnamefont {S.~R.}\ \bibnamefont {Green}}, \bibinfo {author} {\bibfnamefont {S.}~\bibnamefont {Hollands}}, \bibinfo {author} {\bibfnamefont {L.}~\bibnamefont {Sberna}}, \bibinfo {author} {\bibfnamefont {V.}~\bibnamefont {Toomani}}, \ and\ \bibinfo {author} {\bibfnamefont {P.}~\bibnamefont {Zimmerman}},\ }\href {\doibase 10.1103/PhysRevD.107.064030} {\bibfield  {journal} {\bibinfo  {journal} {Phys. Rev. D}\ }\textbf {\bibinfo {volume} {107}},\ \bibinfo {pages} {064030} (\bibinfo {year} {2023})},\ \Eprint {http://arxiv.org/abs/2210.15935} {arXiv:2210.15935 [gr-qc]} \BibitemShut {NoStop}%
\bibitem [{\citenamefont {London}(2023)}]{London:2023aeo}%
  \BibitemOpen
  \bibfield  {author} {\bibinfo {author} {\bibfnamefont {L.}~\bibnamefont {London}},\ }\href@noop {} {\  (\bibinfo {year} {2023})},\ \Eprint {http://arxiv.org/abs/2312.17678} {arXiv:2312.17678 [gr-qc]} \BibitemShut {NoStop}%
\bibitem [{\citenamefont {London}\ and\ \citenamefont {Gurevich}(2023)}]{London:2023idh}%
  \BibitemOpen
  \bibfield  {author} {\bibinfo {author} {\bibfnamefont {L.}~\bibnamefont {London}}\ and\ \bibinfo {author} {\bibfnamefont {M.}~\bibnamefont {Gurevich}},\ }\href@noop {} {\  (\bibinfo {year} {2023})},\ \Eprint {http://arxiv.org/abs/2312.17680} {arXiv:2312.17680 [gr-qc]} \BibitemShut {NoStop}%
\bibitem [{\citenamefont {Leaver}(1985)}]{Leaver:1985ax}%
  \BibitemOpen
  \bibfield  {author} {\bibinfo {author} {\bibfnamefont {E.~W.}\ \bibnamefont {Leaver}},\ }\href {\doibase 10.1098/rspa.1985.0119} {\bibfield  {journal} {\bibinfo  {journal} {Proc. Roy. Soc. Lond. A}\ }\textbf {\bibinfo {volume} {402}},\ \bibinfo {pages} {285} (\bibinfo {year} {1985})}\BibitemShut {NoStop}%
\bibitem [{\citenamefont {Nollert}(1999)}]{Nollert:1999ji}%
  \BibitemOpen
  \bibfield  {author} {\bibinfo {author} {\bibfnamefont {H.-P.}\ \bibnamefont {Nollert}},\ }\href {\doibase 10.1088/0264-9381/16/12/201} {\bibfield  {journal} {\bibinfo  {journal} {Class. Quant. Grav.}\ }\textbf {\bibinfo {volume} {16}},\ \bibinfo {pages} {R159} (\bibinfo {year} {1999})}\BibitemShut {NoStop}%
\bibitem [{\citenamefont {Berti}\ and\ \citenamefont {Cardoso}(2006)}]{berti:2006kk}%
  \BibitemOpen
  \bibfield  {author} {\bibinfo {author} {\bibfnamefont {E.}~\bibnamefont {Berti}}\ and\ \bibinfo {author} {\bibfnamefont {V.}~\bibnamefont {Cardoso}},\ }\href {\doibase 10.1103/PhysRevD.74.104020} {\bibfield  {journal} {\bibinfo  {journal} {Phys. Rev. D}\ }\textbf {\bibinfo {volume} {74}},\ \bibinfo {pages} {104020} (\bibinfo {year} {2006})}\BibitemShut {NoStop}%
\bibitem [{\citenamefont {Albanesi}\ \emph {et~al.}(2023)\citenamefont {Albanesi}, \citenamefont {Bernuzzi}, \citenamefont {Damour}, \citenamefont {Nagar},\ and\ \citenamefont {Placidi}}]{Albanesi:2023bgi}%
  \BibitemOpen
  \bibfield  {author} {\bibinfo {author} {\bibfnamefont {S.}~\bibnamefont {Albanesi}}, \bibinfo {author} {\bibfnamefont {S.}~\bibnamefont {Bernuzzi}}, \bibinfo {author} {\bibfnamefont {T.}~\bibnamefont {Damour}}, \bibinfo {author} {\bibfnamefont {A.}~\bibnamefont {Nagar}}, \ and\ \bibinfo {author} {\bibfnamefont {A.}~\bibnamefont {Placidi}},\ }\href {\doibase 10.1103/PhysRevD.108.084037} {\bibfield  {journal} {\bibinfo  {journal} {Phys. Rev. D}\ }\textbf {\bibinfo {volume} {108}},\ \bibinfo {pages} {084037} (\bibinfo {year} {2023})},\ \Eprint {http://arxiv.org/abs/2305.19336} {arXiv:2305.19336 [gr-qc]} \BibitemShut {NoStop}%
\bibitem [{\citenamefont {Carullo}\ and\ \citenamefont {De~Amicis}(2023)}]{Carullo:2023tff}%
  \BibitemOpen
  \bibfield  {author} {\bibinfo {author} {\bibfnamefont {G.}~\bibnamefont {Carullo}}\ and\ \bibinfo {author} {\bibfnamefont {M.}~\bibnamefont {De~Amicis}},\ }\href@noop {} {\  (\bibinfo {year} {2023})},\ \Eprint {http://arxiv.org/abs/2310.12968} {arXiv:2310.12968 [gr-qc]} \BibitemShut {NoStop}%
\bibitem [{\citenamefont {Baibhav}\ \emph {et~al.}(2023)\citenamefont {Baibhav}, \citenamefont {Cheung}, \citenamefont {Berti}, \citenamefont {Cardoso}, \citenamefont {Carullo}, \citenamefont {Cotesta}, \citenamefont {Del~Pozzo},\ and\ \citenamefont {Duque}}]{Baibhav:2023clw}%
  \BibitemOpen
  \bibfield  {author} {\bibinfo {author} {\bibfnamefont {V.}~\bibnamefont {Baibhav}}, \bibinfo {author} {\bibfnamefont {M.~H.-Y.}\ \bibnamefont {Cheung}}, \bibinfo {author} {\bibfnamefont {E.}~\bibnamefont {Berti}}, \bibinfo {author} {\bibfnamefont {V.}~\bibnamefont {Cardoso}}, \bibinfo {author} {\bibfnamefont {G.}~\bibnamefont {Carullo}}, \bibinfo {author} {\bibfnamefont {R.}~\bibnamefont {Cotesta}}, \bibinfo {author} {\bibfnamefont {W.}~\bibnamefont {Del~Pozzo}}, \ and\ \bibinfo {author} {\bibfnamefont {F.}~\bibnamefont {Duque}},\ }\href@noop {} {\  (\bibinfo {year} {2023})},\ \Eprint {http://arxiv.org/abs/2302.03050} {arXiv:2302.03050 [gr-qc]} \BibitemShut {NoStop}%
\bibitem [{\citenamefont {Zhu}\ \emph {et~al.}(2023{\natexlab{a}})\citenamefont {Zhu}, \citenamefont {Ripley}, \citenamefont {C\'ardenas-Avenda\~no},\ and\ \citenamefont {Pretorius}}]{Zhu:2023mzv}%
  \BibitemOpen
  \bibfield  {author} {\bibinfo {author} {\bibfnamefont {H.}~\bibnamefont {Zhu}}, \bibinfo {author} {\bibfnamefont {J.~L.}\ \bibnamefont {Ripley}}, \bibinfo {author} {\bibfnamefont {A.}~\bibnamefont {C\'ardenas-Avenda\~no}}, \ and\ \bibinfo {author} {\bibfnamefont {F.}~\bibnamefont {Pretorius}},\ }\href@noop {} {\  (\bibinfo {year} {2023}{\natexlab{a}})},\ \Eprint {http://arxiv.org/abs/2309.13204} {arXiv:2309.13204 [gr-qc]} \BibitemShut {NoStop}%
\bibitem [{\citenamefont {Takahashi}\ and\ \citenamefont {Motohashi}(2023)}]{Takahashi:2023tkb}%
  \BibitemOpen
  \bibfield  {author} {\bibinfo {author} {\bibfnamefont {K.}~\bibnamefont {Takahashi}}\ and\ \bibinfo {author} {\bibfnamefont {H.}~\bibnamefont {Motohashi}},\ }\href@noop {} {\  (\bibinfo {year} {2023})},\ \Eprint {http://arxiv.org/abs/2311.12762} {arXiv:2311.12762 [gr-qc]} \BibitemShut {NoStop}%
\bibitem [{\citenamefont {Gleiser}\ \emph {et~al.}(1996{\natexlab{a}})\citenamefont {Gleiser}, \citenamefont {Nicasio}, \citenamefont {Price},\ and\ \citenamefont {Pullin}}]{Gleiser:1995gx}%
  \BibitemOpen
  \bibfield  {author} {\bibinfo {author} {\bibfnamefont {R.~J.}\ \bibnamefont {Gleiser}}, \bibinfo {author} {\bibfnamefont {C.~O.}\ \bibnamefont {Nicasio}}, \bibinfo {author} {\bibfnamefont {R.~H.}\ \bibnamefont {Price}}, \ and\ \bibinfo {author} {\bibfnamefont {J.}~\bibnamefont {Pullin}},\ }\href {\doibase 10.1088/0264-9381/13/10/001} {\bibfield  {journal} {\bibinfo  {journal} {Class. Quant. Grav.}\ }\textbf {\bibinfo {volume} {13}},\ \bibinfo {pages} {L117} (\bibinfo {year} {1996}{\natexlab{a}})},\ \Eprint {http://arxiv.org/abs/gr-qc/9510049} {arXiv:gr-qc/9510049} \BibitemShut {NoStop}%
\bibitem [{\citenamefont {Brizuela}\ \emph {et~al.}(2009)\citenamefont {Brizuela}, \citenamefont {Martin-Garcia},\ and\ \citenamefont {Tiglio}}]{Brizuela:2009qd}%
  \BibitemOpen
  \bibfield  {author} {\bibinfo {author} {\bibfnamefont {D.}~\bibnamefont {Brizuela}}, \bibinfo {author} {\bibfnamefont {J.~M.}\ \bibnamefont {Martin-Garcia}}, \ and\ \bibinfo {author} {\bibfnamefont {M.}~\bibnamefont {Tiglio}},\ }\href {\doibase 10.1103/PhysRevD.80.024021} {\bibfield  {journal} {\bibinfo  {journal} {Phys. Rev. D}\ }\textbf {\bibinfo {volume} {80}},\ \bibinfo {pages} {024021} (\bibinfo {year} {2009})},\ \Eprint {http://arxiv.org/abs/0903.1134} {arXiv:0903.1134 [gr-qc]} \BibitemShut {NoStop}%
\bibitem [{\citenamefont {Ripley}\ \emph {et~al.}(2021)\citenamefont {Ripley}, \citenamefont {Loutrel}, \citenamefont {Giorgi},\ and\ \citenamefont {Pretorius}}]{Ripley:2020xby}%
  \BibitemOpen
  \bibfield  {author} {\bibinfo {author} {\bibfnamefont {J.~L.}\ \bibnamefont {Ripley}}, \bibinfo {author} {\bibfnamefont {N.}~\bibnamefont {Loutrel}}, \bibinfo {author} {\bibfnamefont {E.}~\bibnamefont {Giorgi}}, \ and\ \bibinfo {author} {\bibfnamefont {F.}~\bibnamefont {Pretorius}},\ }\href {\doibase 10.1103/PhysRevD.103.104018} {\bibfield  {journal} {\bibinfo  {journal} {Phys. Rev. D}\ }\textbf {\bibinfo {volume} {103}},\ \bibinfo {pages} {104018} (\bibinfo {year} {2021})},\ \Eprint {http://arxiv.org/abs/2010.00162} {arXiv:2010.00162 [gr-qc]} \BibitemShut {NoStop}%
\bibitem [{\citenamefont {Ioka}\ and\ \citenamefont {Nakano}(2007)}]{Ioka:2007ak}%
  \BibitemOpen
  \bibfield  {author} {\bibinfo {author} {\bibfnamefont {K.}~\bibnamefont {Ioka}}\ and\ \bibinfo {author} {\bibfnamefont {H.}~\bibnamefont {Nakano}},\ }\href {\doibase 10.1103/PhysRevD.76.061503} {\bibfield  {journal} {\bibinfo  {journal} {Phys. Rev. D}\ }\textbf {\bibinfo {volume} {76}},\ \bibinfo {pages} {061503} (\bibinfo {year} {2007})},\ \Eprint {http://arxiv.org/abs/0704.3467} {arXiv:0704.3467 [astro-ph]} \BibitemShut {NoStop}%
\bibitem [{\citenamefont {Nakano}\ and\ \citenamefont {Ioka}(2007)}]{Nakano:2007cj}%
  \BibitemOpen
  \bibfield  {author} {\bibinfo {author} {\bibfnamefont {H.}~\bibnamefont {Nakano}}\ and\ \bibinfo {author} {\bibfnamefont {K.}~\bibnamefont {Ioka}},\ }\href {\doibase 10.1103/PhysRevD.76.084007} {\bibfield  {journal} {\bibinfo  {journal} {Phys. Rev. D}\ }\textbf {\bibinfo {volume} {76}},\ \bibinfo {pages} {084007} (\bibinfo {year} {2007})},\ \Eprint {http://arxiv.org/abs/0708.0450} {arXiv:0708.0450 [gr-qc]} \BibitemShut {NoStop}%
\bibitem [{\citenamefont {Pazos}\ \emph {et~al.}(2010)\citenamefont {Pazos}, \citenamefont {Brizuela}, \citenamefont {Martin-Garcia},\ and\ \citenamefont {Tiglio}}]{Pazos:2010xf}%
  \BibitemOpen
  \bibfield  {author} {\bibinfo {author} {\bibfnamefont {E.}~\bibnamefont {Pazos}}, \bibinfo {author} {\bibfnamefont {D.}~\bibnamefont {Brizuela}}, \bibinfo {author} {\bibfnamefont {J.~M.}\ \bibnamefont {Martin-Garcia}}, \ and\ \bibinfo {author} {\bibfnamefont {M.}~\bibnamefont {Tiglio}},\ }\href {\doibase 10.1103/PhysRevD.82.104028} {\bibfield  {journal} {\bibinfo  {journal} {Phys. Rev. D}\ }\textbf {\bibinfo {volume} {82}},\ \bibinfo {pages} {104028} (\bibinfo {year} {2010})},\ \Eprint {http://arxiv.org/abs/1009.4665} {arXiv:1009.4665 [gr-qc]} \BibitemShut {NoStop}%
\bibitem [{\citenamefont {London}\ \emph {et~al.}(2014)\citenamefont {London}, \citenamefont {Shoemaker},\ and\ \citenamefont {Healy}}]{London:2014cma}%
  \BibitemOpen
  \bibfield  {author} {\bibinfo {author} {\bibfnamefont {L.}~\bibnamefont {London}}, \bibinfo {author} {\bibfnamefont {D.}~\bibnamefont {Shoemaker}}, \ and\ \bibinfo {author} {\bibfnamefont {J.}~\bibnamefont {Healy}},\ }\href {\doibase 10.1103/PhysRevD.90.124032} {\bibfield  {journal} {\bibinfo  {journal} {Phys. Rev. D}\ }\textbf {\bibinfo {volume} {90}},\ \bibinfo {pages} {124032} (\bibinfo {year} {2014})},\ \bibinfo {note} {[Erratum: Phys.Rev.D 94, 069902 (2016)]},\ \Eprint {http://arxiv.org/abs/1404.3197} {arXiv:1404.3197 [gr-qc]} \BibitemShut {NoStop}%
\bibitem [{\citenamefont {Lagos}\ and\ \citenamefont {Hui}(2023)}]{Lagos:2022otp}%
  \BibitemOpen
  \bibfield  {author} {\bibinfo {author} {\bibfnamefont {M.}~\bibnamefont {Lagos}}\ and\ \bibinfo {author} {\bibfnamefont {L.}~\bibnamefont {Hui}},\ }\href {\doibase 10.1103/PhysRevD.107.044040} {\bibfield  {journal} {\bibinfo  {journal} {Phys. Rev. D}\ }\textbf {\bibinfo {volume} {107}},\ \bibinfo {pages} {044040} (\bibinfo {year} {2023})},\ \Eprint {http://arxiv.org/abs/2208.07379} {arXiv:2208.07379 [gr-qc]} \BibitemShut {NoStop}%
\bibitem [{\citenamefont {Khera}\ \emph {et~al.}(2023)\citenamefont {Khera}, \citenamefont {Ribes~Metidieri}, \citenamefont {Bonga}, \citenamefont {Forteza}, \citenamefont {Krishnan}, \citenamefont {Poisson}, \citenamefont {Pook-Kolb}, \citenamefont {Schnetter},\ and\ \citenamefont {Yang}}]{Khera:2023lnc}%
  \BibitemOpen
  \bibfield  {author} {\bibinfo {author} {\bibfnamefont {N.}~\bibnamefont {Khera}}, \bibinfo {author} {\bibfnamefont {A.}~\bibnamefont {Ribes~Metidieri}}, \bibinfo {author} {\bibfnamefont {B.}~\bibnamefont {Bonga}}, \bibinfo {author} {\bibfnamefont {X.~J.}\ \bibnamefont {Forteza}}, \bibinfo {author} {\bibfnamefont {B.}~\bibnamefont {Krishnan}}, \bibinfo {author} {\bibfnamefont {E.}~\bibnamefont {Poisson}}, \bibinfo {author} {\bibfnamefont {D.}~\bibnamefont {Pook-Kolb}}, \bibinfo {author} {\bibfnamefont {E.}~\bibnamefont {Schnetter}}, \ and\ \bibinfo {author} {\bibfnamefont {H.}~\bibnamefont {Yang}},\ }\href@noop {} {\  (\bibinfo {year} {2023})},\ \Eprint {http://arxiv.org/abs/2306.11142} {arXiv:2306.11142 [gr-qc]} \BibitemShut {NoStop}%
\bibitem [{\citenamefont {Redondo-Yuste}\ \emph {et~al.}(2023)\citenamefont {Redondo-Yuste}, \citenamefont {Carullo}, \citenamefont {Ripley}, \citenamefont {Berti},\ and\ \citenamefont {Cardoso}}]{Redondo-Yuste:2023seq}%
  \BibitemOpen
  \bibfield  {author} {\bibinfo {author} {\bibfnamefont {J.}~\bibnamefont {Redondo-Yuste}}, \bibinfo {author} {\bibfnamefont {G.}~\bibnamefont {Carullo}}, \bibinfo {author} {\bibfnamefont {J.~L.}\ \bibnamefont {Ripley}}, \bibinfo {author} {\bibfnamefont {E.}~\bibnamefont {Berti}}, \ and\ \bibinfo {author} {\bibfnamefont {V.}~\bibnamefont {Cardoso}},\ }\href@noop {} {\  (\bibinfo {year} {2023})},\ \Eprint {http://arxiv.org/abs/2308.14796} {arXiv:2308.14796 [gr-qc]} \BibitemShut {NoStop}%
\bibitem [{\citenamefont {Zhu}\ \emph {et~al.}(2024)\citenamefont {Zhu} \emph {et~al.}}]{Zhu:2024rej}%
  \BibitemOpen
  \bibfield  {author} {\bibinfo {author} {\bibfnamefont {H.}~\bibnamefont {Zhu}} \emph {et~al.},\ }\href@noop {} {\  (\bibinfo {year} {2024})},\ \Eprint {http://arxiv.org/abs/2401.00805} {arXiv:2401.00805 [gr-qc]} \BibitemShut {NoStop}%
\bibitem [{\citenamefont {Ma}\ and\ \citenamefont {Yang}(2024)}]{Ma:2024qcv}%
  \BibitemOpen
  \bibfield  {author} {\bibinfo {author} {\bibfnamefont {S.}~\bibnamefont {Ma}}\ and\ \bibinfo {author} {\bibfnamefont {H.}~\bibnamefont {Yang}},\ }\href@noop {} {\  (\bibinfo {year} {2024})},\ \Eprint {http://arxiv.org/abs/2401.15516} {arXiv:2401.15516 [gr-qc]} \BibitemShut {NoStop}%
\bibitem [{\citenamefont {Zlochower}\ \emph {et~al.}(2003)\citenamefont {Zlochower}, \citenamefont {Gomez}, \citenamefont {Husa}, \citenamefont {Lehner},\ and\ \citenamefont {Winicour}}]{Zlochower:2003yh}%
  \BibitemOpen
  \bibfield  {author} {\bibinfo {author} {\bibfnamefont {Y.}~\bibnamefont {Zlochower}}, \bibinfo {author} {\bibfnamefont {R.}~\bibnamefont {Gomez}}, \bibinfo {author} {\bibfnamefont {S.}~\bibnamefont {Husa}}, \bibinfo {author} {\bibfnamefont {L.}~\bibnamefont {Lehner}}, \ and\ \bibinfo {author} {\bibfnamefont {J.}~\bibnamefont {Winicour}},\ }\href {\doibase 10.1103/PhysRevD.68.084014} {\bibfield  {journal} {\bibinfo  {journal} {Phys. Rev. D}\ }\textbf {\bibinfo {volume} {68}},\ \bibinfo {pages} {084014} (\bibinfo {year} {2003})},\ \Eprint {http://arxiv.org/abs/gr-qc/0306098} {arXiv:gr-qc/0306098} \BibitemShut {NoStop}%
\bibitem [{\citenamefont {Mitman}\ \emph {et~al.}(2023)\citenamefont {Mitman} \emph {et~al.}}]{Mitman:2022qdl}%
  \BibitemOpen
  \bibfield  {author} {\bibinfo {author} {\bibfnamefont {K.}~\bibnamefont {Mitman}} \emph {et~al.},\ }\href {\doibase 10.1103/PhysRevLett.130.081402} {\bibfield  {journal} {\bibinfo  {journal} {Phys. Rev. Lett.}\ }\textbf {\bibinfo {volume} {130}},\ \bibinfo {pages} {081402} (\bibinfo {year} {2023})},\ \Eprint {http://arxiv.org/abs/2208.07380} {arXiv:2208.07380 [gr-qc]} \BibitemShut {NoStop}%
\bibitem [{\citenamefont {Cheung}\ \emph {et~al.}(2023{\natexlab{a}})\citenamefont {Cheung} \emph {et~al.}}]{Cheung:2022rbm}%
  \BibitemOpen
  \bibfield  {author} {\bibinfo {author} {\bibfnamefont {M.~H.-Y.}\ \bibnamefont {Cheung}} \emph {et~al.},\ }\href {\doibase 10.1103/PhysRevLett.130.081401} {\bibfield  {journal} {\bibinfo  {journal} {Phys. Rev. Lett.}\ }\textbf {\bibinfo {volume} {130}},\ \bibinfo {pages} {081401} (\bibinfo {year} {2023}{\natexlab{a}})},\ \Eprint {http://arxiv.org/abs/2208.07374} {arXiv:2208.07374 [gr-qc]} \BibitemShut {NoStop}%
\bibitem [{\citenamefont {Cheung}\ \emph {et~al.}(2023{\natexlab{b}})\citenamefont {Cheung}, \citenamefont {Berti}, \citenamefont {Baibhav},\ and\ \citenamefont {Cotesta}}]{cheung2023extracting}%
  \BibitemOpen
  \bibfield  {author} {\bibinfo {author} {\bibfnamefont {M.~H.-Y.}\ \bibnamefont {Cheung}}, \bibinfo {author} {\bibfnamefont {E.}~\bibnamefont {Berti}}, \bibinfo {author} {\bibfnamefont {V.}~\bibnamefont {Baibhav}}, \ and\ \bibinfo {author} {\bibfnamefont {R.}~\bibnamefont {Cotesta}},\ }\href@noop {} {\enquote {\bibinfo {title} {Extracting linear and nonlinear quasinormal modes from black hole merger simulations},}\ } (\bibinfo {year} {2023}{\natexlab{b}}),\ \Eprint {http://arxiv.org/abs/2310.04489} {arXiv:2310.04489 [gr-qc]} \BibitemShut {NoStop}%
\bibitem [{\citenamefont {Ma}\ \emph {et~al.}(2022)\citenamefont {Ma}, \citenamefont {Mitman}, \citenamefont {Sun}, \citenamefont {Deppe}, \citenamefont {H\'ebert}, \citenamefont {Kidder}, \citenamefont {Moxon}, \citenamefont {Throwe}, \citenamefont {Vu},\ and\ \citenamefont {Chen}}]{Ma:2022wpv}%
  \BibitemOpen
  \bibfield  {author} {\bibinfo {author} {\bibfnamefont {S.}~\bibnamefont {Ma}}, \bibinfo {author} {\bibfnamefont {K.}~\bibnamefont {Mitman}}, \bibinfo {author} {\bibfnamefont {L.}~\bibnamefont {Sun}}, \bibinfo {author} {\bibfnamefont {N.}~\bibnamefont {Deppe}}, \bibinfo {author} {\bibfnamefont {F.}~\bibnamefont {H\'ebert}}, \bibinfo {author} {\bibfnamefont {L.~E.}\ \bibnamefont {Kidder}}, \bibinfo {author} {\bibfnamefont {J.}~\bibnamefont {Moxon}}, \bibinfo {author} {\bibfnamefont {W.}~\bibnamefont {Throwe}}, \bibinfo {author} {\bibfnamefont {N.~L.}\ \bibnamefont {Vu}}, \ and\ \bibinfo {author} {\bibfnamefont {Y.}~\bibnamefont {Chen}},\ }\href {\doibase 10.1103/PhysRevD.106.084036} {\bibfield  {journal} {\bibinfo  {journal} {Phys. Rev. D}\ }\textbf {\bibinfo {volume} {106}},\ \bibinfo {pages} {084036} (\bibinfo {year} {2022})},\ \Eprint {http://arxiv.org/abs/2207.10870} {arXiv:2207.10870 [gr-qc]} \BibitemShut {NoStop}%
\bibitem [{\citenamefont {Yi}\ \emph {et~al.}(2024)\citenamefont {Yi}, \citenamefont {Kuntz}, \citenamefont {Barausse}, \citenamefont {Berti}, \citenamefont {Cheung}, \citenamefont {Kritos},\ and\ \citenamefont {Maselli}}]{Yi:2024elj}%
  \BibitemOpen
  \bibfield  {author} {\bibinfo {author} {\bibfnamefont {S.}~\bibnamefont {Yi}}, \bibinfo {author} {\bibfnamefont {A.}~\bibnamefont {Kuntz}}, \bibinfo {author} {\bibfnamefont {E.}~\bibnamefont {Barausse}}, \bibinfo {author} {\bibfnamefont {E.}~\bibnamefont {Berti}}, \bibinfo {author} {\bibfnamefont {M.~H.-Y.}\ \bibnamefont {Cheung}}, \bibinfo {author} {\bibfnamefont {K.}~\bibnamefont {Kritos}}, \ and\ \bibinfo {author} {\bibfnamefont {A.}~\bibnamefont {Maselli}},\ }\href@noop {} {\  (\bibinfo {year} {2024})},\ \Eprint {http://arxiv.org/abs/2403.09767} {arXiv:2403.09767 [gr-qc]} \BibitemShut {NoStop}%
\bibitem [{\citenamefont {Sberna}\ \emph {et~al.}(2022)\citenamefont {Sberna}, \citenamefont {Bosch}, \citenamefont {East}, \citenamefont {Green},\ and\ \citenamefont {Lehner}}]{Sberna:2021eui}%
  \BibitemOpen
  \bibfield  {author} {\bibinfo {author} {\bibfnamefont {L.}~\bibnamefont {Sberna}}, \bibinfo {author} {\bibfnamefont {P.}~\bibnamefont {Bosch}}, \bibinfo {author} {\bibfnamefont {W.~E.}\ \bibnamefont {East}}, \bibinfo {author} {\bibfnamefont {S.~R.}\ \bibnamefont {Green}}, \ and\ \bibinfo {author} {\bibfnamefont {L.}~\bibnamefont {Lehner}},\ }\href {\doibase 10.1103/PhysRevD.105.064046} {\bibfield  {journal} {\bibinfo  {journal} {Phys. Rev. D}\ }\textbf {\bibinfo {volume} {105}},\ \bibinfo {pages} {064046} (\bibinfo {year} {2022})},\ \Eprint {http://arxiv.org/abs/2112.11168} {arXiv:2112.11168 [gr-qc]} \BibitemShut {NoStop}%
\bibitem [{\citenamefont {Cannizzaro}\ \emph {et~al.}(2024)\citenamefont {Cannizzaro}, \citenamefont {Sberna}, \citenamefont {Green},\ and\ \citenamefont {Hollands}}]{Cannizzaro:2023jle}%
  \BibitemOpen
  \bibfield  {author} {\bibinfo {author} {\bibfnamefont {E.}~\bibnamefont {Cannizzaro}}, \bibinfo {author} {\bibfnamefont {L.}~\bibnamefont {Sberna}}, \bibinfo {author} {\bibfnamefont {S.~R.}\ \bibnamefont {Green}}, \ and\ \bibinfo {author} {\bibfnamefont {S.}~\bibnamefont {Hollands}},\ }\href {\doibase 10.1103/PhysRevLett.132.051401} {\bibfield  {journal} {\bibinfo  {journal} {Phys. Rev. Lett.}\ }\textbf {\bibinfo {volume} {132}},\ \bibinfo {pages} {051401} (\bibinfo {year} {2024})},\ \Eprint {http://arxiv.org/abs/2309.10021} {arXiv:2309.10021 [gr-qc]} \BibitemShut {NoStop}%
\bibitem [{\citenamefont {Redondo-Yuste}\ \emph {et~al.}(2024)\citenamefont {Redondo-Yuste}, \citenamefont {Pere\~niguez},\ and\ \citenamefont {Cardoso}}]{Redondo-Yuste:2023ipg}%
  \BibitemOpen
  \bibfield  {author} {\bibinfo {author} {\bibfnamefont {J.}~\bibnamefont {Redondo-Yuste}}, \bibinfo {author} {\bibfnamefont {D.}~\bibnamefont {Pere\~niguez}}, \ and\ \bibinfo {author} {\bibfnamefont {V.}~\bibnamefont {Cardoso}},\ }\href {\doibase 10.1103/PhysRevD.109.044048} {\bibfield  {journal} {\bibinfo  {journal} {Phys. Rev. D}\ }\textbf {\bibinfo {volume} {109}},\ \bibinfo {pages} {044048} (\bibinfo {year} {2024})},\ \Eprint {http://arxiv.org/abs/2312.04633} {arXiv:2312.04633 [gr-qc]} \BibitemShut {NoStop}%
\bibitem [{\citenamefont {Pretorius}(2005{\natexlab{a}})}]{Pretorius:2005gq}%
  \BibitemOpen
  \bibfield  {author} {\bibinfo {author} {\bibfnamefont {F.}~\bibnamefont {Pretorius}},\ }\href {\doibase 10.1103/PhysRevLett.95.121101} {\bibfield  {journal} {\bibinfo  {journal} {Phys. Rev. Lett.}\ }\textbf {\bibinfo {volume} {95}},\ \bibinfo {pages} {121101} (\bibinfo {year} {2005}{\natexlab{a}})},\ \Eprint {http://arxiv.org/abs/gr-qc/0507014} {arXiv:gr-qc/0507014} \BibitemShut {NoStop}%
\bibitem [{\citenamefont {Green}\ \emph {et~al.}(2020)\citenamefont {Green}, \citenamefont {Hollands},\ and\ \citenamefont {Zimmerman}}]{Green:2019nam}%
  \BibitemOpen
  \bibfield  {author} {\bibinfo {author} {\bibfnamefont {S.~R.}\ \bibnamefont {Green}}, \bibinfo {author} {\bibfnamefont {S.}~\bibnamefont {Hollands}}, \ and\ \bibinfo {author} {\bibfnamefont {P.}~\bibnamefont {Zimmerman}},\ }\href {\doibase 10.1088/1361-6382/ab7075} {\bibfield  {journal} {\bibinfo  {journal} {Class. Quant. Grav.}\ }\textbf {\bibinfo {volume} {37}},\ \bibinfo {pages} {075001} (\bibinfo {year} {2020})},\ \Eprint {http://arxiv.org/abs/1908.09095} {arXiv:1908.09095 [gr-qc]} \BibitemShut {NoStop}%
\bibitem [{\citenamefont {Andersson}\ \emph {et~al.}(2022)\citenamefont {Andersson}, \citenamefont {B\"ackdahl}, \citenamefont {Blue},\ and\ \citenamefont {Ma}}]{Andersson:2021eqc}%
  \BibitemOpen
  \bibfield  {author} {\bibinfo {author} {\bibfnamefont {L.}~\bibnamefont {Andersson}}, \bibinfo {author} {\bibfnamefont {T.}~\bibnamefont {B\"ackdahl}}, \bibinfo {author} {\bibfnamefont {P.}~\bibnamefont {Blue}}, \ and\ \bibinfo {author} {\bibfnamefont {S.}~\bibnamefont {Ma}},\ }\href {\doibase 10.1007/s00220-022-04461-3} {\bibfield  {journal} {\bibinfo  {journal} {Commun. Math. Phys.}\ }\textbf {\bibinfo {volume} {396}},\ \bibinfo {pages} {45} (\bibinfo {year} {2022})},\ \Eprint {http://arxiv.org/abs/2108.03148} {arXiv:2108.03148 [gr-qc]} \BibitemShut {NoStop}%
\bibitem [{\citenamefont {Loutrel}\ \emph {et~al.}(2021)\citenamefont {Loutrel}, \citenamefont {Ripley}, \citenamefont {Giorgi},\ and\ \citenamefont {Pretorius}}]{Loutrel:2020wbw}%
  \BibitemOpen
  \bibfield  {author} {\bibinfo {author} {\bibfnamefont {N.}~\bibnamefont {Loutrel}}, \bibinfo {author} {\bibfnamefont {J.~L.}\ \bibnamefont {Ripley}}, \bibinfo {author} {\bibfnamefont {E.}~\bibnamefont {Giorgi}}, \ and\ \bibinfo {author} {\bibfnamefont {F.}~\bibnamefont {Pretorius}},\ }\href {\doibase 10.1103/PhysRevD.103.104017} {\bibfield  {journal} {\bibinfo  {journal} {Phys. Rev. D}\ }\textbf {\bibinfo {volume} {103}},\ \bibinfo {pages} {104017} (\bibinfo {year} {2021})},\ \Eprint {http://arxiv.org/abs/2008.11770} {arXiv:2008.11770 [gr-qc]} \BibitemShut {NoStop}%
\bibitem [{\citenamefont {Toomani}\ \emph {et~al.}(2022)\citenamefont {Toomani}, \citenamefont {Zimmerman}, \citenamefont {Spiers}, \citenamefont {Hollands}, \citenamefont {Pound},\ and\ \citenamefont {Green}}]{Toomani:2021jlo}%
  \BibitemOpen
  \bibfield  {author} {\bibinfo {author} {\bibfnamefont {V.}~\bibnamefont {Toomani}}, \bibinfo {author} {\bibfnamefont {P.}~\bibnamefont {Zimmerman}}, \bibinfo {author} {\bibfnamefont {A.}~\bibnamefont {Spiers}}, \bibinfo {author} {\bibfnamefont {S.}~\bibnamefont {Hollands}}, \bibinfo {author} {\bibfnamefont {A.}~\bibnamefont {Pound}}, \ and\ \bibinfo {author} {\bibfnamefont {S.~R.}\ \bibnamefont {Green}},\ }\href {\doibase 10.1088/1361-6382/ac37a5} {\bibfield  {journal} {\bibinfo  {journal} {Class. Quant. Grav.}\ }\textbf {\bibinfo {volume} {39}},\ \bibinfo {pages} {015019} (\bibinfo {year} {2022})},\ \Eprint {http://arxiv.org/abs/2108.04273} {arXiv:2108.04273 [gr-qc]} \BibitemShut {NoStop}%
\bibitem [{\citenamefont {Spiers}\ \emph {et~al.}(2023)\citenamefont {Spiers}, \citenamefont {Pound},\ and\ \citenamefont {Moxon}}]{Spiers:2023cip}%
  \BibitemOpen
  \bibfield  {author} {\bibinfo {author} {\bibfnamefont {A.}~\bibnamefont {Spiers}}, \bibinfo {author} {\bibfnamefont {A.}~\bibnamefont {Pound}}, \ and\ \bibinfo {author} {\bibfnamefont {J.}~\bibnamefont {Moxon}},\ }\href {\doibase 10.1103/PhysRevD.108.064002} {\bibfield  {journal} {\bibinfo  {journal} {Phys. Rev. D}\ }\textbf {\bibinfo {volume} {108}},\ \bibinfo {pages} {064002} (\bibinfo {year} {2023})},\ \Eprint {http://arxiv.org/abs/2305.19332} {arXiv:2305.19332 [gr-qc]} \BibitemShut {NoStop}%
\bibitem [{\citenamefont {May}\ \emph {et~al.}(2023)\citenamefont {May}, \citenamefont {Ripley},\ and\ \citenamefont {East}}]{May_in_prep}%
  \BibitemOpen
  \bibfield  {author} {\bibinfo {author} {\bibfnamefont {T.}~\bibnamefont {May}}, \bibinfo {author} {\bibfnamefont {J.}~\bibnamefont {Ripley}}, \ and\ \bibinfo {author} {\bibfnamefont {W.}~\bibnamefont {East}},\ }\href@noop {} {\bibfield  {journal} {\bibinfo  {journal} {In Preperation}\ } (\bibinfo {year} {2023})}\BibitemShut {NoStop}%
\bibitem [{\citenamefont {Price}\ and\ \citenamefont {Pullin}(1994)}]{Price:1994pm}%
  \BibitemOpen
  \bibfield  {author} {\bibinfo {author} {\bibfnamefont {R.~H.}\ \bibnamefont {Price}}\ and\ \bibinfo {author} {\bibfnamefont {J.}~\bibnamefont {Pullin}},\ }\href {\doibase 10.1103/PhysRevLett.72.3297} {\bibfield  {journal} {\bibinfo  {journal} {Phys. Rev. Lett.}\ }\textbf {\bibinfo {volume} {72}},\ \bibinfo {pages} {3297} (\bibinfo {year} {1994})},\ \Eprint {http://arxiv.org/abs/gr-qc/9402039} {arXiv:gr-qc/9402039} \BibitemShut {NoStop}%
\bibitem [{\citenamefont {Gleiser}\ \emph {et~al.}(1996{\natexlab{b}})\citenamefont {Gleiser}, \citenamefont {Nicasio}, \citenamefont {Price},\ and\ \citenamefont {Pullin}}]{Gleiser:1996yc}%
  \BibitemOpen
  \bibfield  {author} {\bibinfo {author} {\bibfnamefont {R.~J.}\ \bibnamefont {Gleiser}}, \bibinfo {author} {\bibfnamefont {C.~O.}\ \bibnamefont {Nicasio}}, \bibinfo {author} {\bibfnamefont {R.~H.}\ \bibnamefont {Price}}, \ and\ \bibinfo {author} {\bibfnamefont {J.}~\bibnamefont {Pullin}},\ }\href {\doibase 10.1103/PhysRevLett.77.4483} {\bibfield  {journal} {\bibinfo  {journal} {Phys. Rev. Lett.}\ }\textbf {\bibinfo {volume} {77}},\ \bibinfo {pages} {4483} (\bibinfo {year} {1996}{\natexlab{b}})},\ \Eprint {http://arxiv.org/abs/gr-qc/9609022} {arXiv:gr-qc/9609022} \BibitemShut {NoStop}%
\bibitem [{\citenamefont {Andrade}\ and\ \citenamefont {Price}(1997)}]{Andrade:1996pc}%
  \BibitemOpen
  \bibfield  {author} {\bibinfo {author} {\bibfnamefont {Z.}~\bibnamefont {Andrade}}\ and\ \bibinfo {author} {\bibfnamefont {R.~H.}\ \bibnamefont {Price}},\ }\href {\doibase 10.1103/PhysRevD.56.6336} {\bibfield  {journal} {\bibinfo  {journal} {Phys. Rev. D}\ }\textbf {\bibinfo {volume} {56}},\ \bibinfo {pages} {6336} (\bibinfo {year} {1997})},\ \Eprint {http://arxiv.org/abs/gr-qc/9611022} {arXiv:gr-qc/9611022} \BibitemShut {NoStop}%
\bibitem [{\citenamefont {Khanna}\ \emph {et~al.}(1999)\citenamefont {Khanna}, \citenamefont {Baker}, \citenamefont {Gleiser}, \citenamefont {Laguna}, \citenamefont {Nicasio}, \citenamefont {Nollert}, \citenamefont {Price},\ and\ \citenamefont {Pullin}}]{Khanna:1999mh}%
  \BibitemOpen
  \bibfield  {author} {\bibinfo {author} {\bibfnamefont {G.}~\bibnamefont {Khanna}}, \bibinfo {author} {\bibfnamefont {J.~G.}\ \bibnamefont {Baker}}, \bibinfo {author} {\bibfnamefont {R.~J.}\ \bibnamefont {Gleiser}}, \bibinfo {author} {\bibfnamefont {P.}~\bibnamefont {Laguna}}, \bibinfo {author} {\bibfnamefont {C.~O.}\ \bibnamefont {Nicasio}}, \bibinfo {author} {\bibfnamefont {H.-P.}\ \bibnamefont {Nollert}}, \bibinfo {author} {\bibfnamefont {R.}~\bibnamefont {Price}}, \ and\ \bibinfo {author} {\bibfnamefont {J.}~\bibnamefont {Pullin}},\ }\href {\doibase 10.1103/PhysRevLett.83.3581} {\bibfield  {journal} {\bibinfo  {journal} {Phys. Rev. Lett.}\ }\textbf {\bibinfo {volume} {83}},\ \bibinfo {pages} {3581} (\bibinfo {year} {1999})},\ \Eprint {http://arxiv.org/abs/gr-qc/9905081} {arXiv:gr-qc/9905081} \BibitemShut {NoStop}%
\bibitem [{\citenamefont {Allen}\ \emph {et~al.}(1998)\citenamefont {Allen}, \citenamefont {Camarda},\ and\ \citenamefont {Seidel}}]{Allen:1998rg}%
  \BibitemOpen
  \bibfield  {author} {\bibinfo {author} {\bibfnamefont {G.}~\bibnamefont {Allen}}, \bibinfo {author} {\bibfnamefont {K.}~\bibnamefont {Camarda}}, \ and\ \bibinfo {author} {\bibfnamefont {E.}~\bibnamefont {Seidel}},\ }\href@noop {} {\  (\bibinfo {year} {1998})},\ \Eprint {http://arxiv.org/abs/gr-qc/9806014} {arXiv:gr-qc/9806014} \BibitemShut {NoStop}%
\bibitem [{\citenamefont {Baker}\ \emph {et~al.}(2000)\citenamefont {Baker}, \citenamefont {Brandt}, \citenamefont {Campanelli}, \citenamefont {Lousto}, \citenamefont {Seidel},\ and\ \citenamefont {Takahashi}}]{Baker:1999sj}%
  \BibitemOpen
  \bibfield  {author} {\bibinfo {author} {\bibfnamefont {J.~G.}\ \bibnamefont {Baker}}, \bibinfo {author} {\bibfnamefont {S.}~\bibnamefont {Brandt}}, \bibinfo {author} {\bibfnamefont {M.}~\bibnamefont {Campanelli}}, \bibinfo {author} {\bibfnamefont {C.~O.}\ \bibnamefont {Lousto}}, \bibinfo {author} {\bibfnamefont {E.}~\bibnamefont {Seidel}}, \ and\ \bibinfo {author} {\bibfnamefont {R.}~\bibnamefont {Takahashi}},\ }\href {\doibase 10.1103/PhysRevD.62.127701} {\bibfield  {journal} {\bibinfo  {journal} {Phys. Rev. D}\ }\textbf {\bibinfo {volume} {62}},\ \bibinfo {pages} {127701} (\bibinfo {year} {2000})},\ \Eprint {http://arxiv.org/abs/gr-qc/9911017} {arXiv:gr-qc/9911017} \BibitemShut {NoStop}%
\bibitem [{\citenamefont {Tichy}\ and\ \citenamefont {Marronetti}(2008)}]{Tichy:2008du}%
  \BibitemOpen
  \bibfield  {author} {\bibinfo {author} {\bibfnamefont {W.}~\bibnamefont {Tichy}}\ and\ \bibinfo {author} {\bibfnamefont {P.}~\bibnamefont {Marronetti}},\ }\href {\doibase 10.1103/PhysRevD.78.081501} {\bibfield  {journal} {\bibinfo  {journal} {Phys. Rev. D}\ }\textbf {\bibinfo {volume} {78}},\ \bibinfo {pages} {081501} (\bibinfo {year} {2008})},\ \Eprint {http://arxiv.org/abs/0807.2985} {arXiv:0807.2985 [gr-qc]} \BibitemShut {NoStop}%
\bibitem [{\citenamefont {Isi}\ and\ \citenamefont {Farr}(2021)}]{Isi:2021iql}%
  \BibitemOpen
  \bibfield  {author} {\bibinfo {author} {\bibfnamefont {M.}~\bibnamefont {Isi}}\ and\ \bibinfo {author} {\bibfnamefont {W.~M.}\ \bibnamefont {Farr}},\ }\href@noop {} {\  (\bibinfo {year} {2021})},\ \Eprint {http://arxiv.org/abs/2107.05609} {arXiv:2107.05609 [gr-qc]} \BibitemShut {NoStop}%
\bibitem [{\citenamefont {Pfeiffer}\ \emph {et~al.}(2005)\citenamefont {Pfeiffer}, \citenamefont {Kidder}, \citenamefont {Scheel},\ and\ \citenamefont {Shoemaker}}]{Pfeiffer:2004qz}%
  \BibitemOpen
  \bibfield  {author} {\bibinfo {author} {\bibfnamefont {H.~P.}\ \bibnamefont {Pfeiffer}}, \bibinfo {author} {\bibfnamefont {L.~E.}\ \bibnamefont {Kidder}}, \bibinfo {author} {\bibfnamefont {M.~A.}\ \bibnamefont {Scheel}}, \ and\ \bibinfo {author} {\bibfnamefont {D.}~\bibnamefont {Shoemaker}},\ }\href {\doibase 10.1103/PhysRevD.71.024020} {\bibfield  {journal} {\bibinfo  {journal} {Phys. Rev. D}\ }\textbf {\bibinfo {volume} {71}},\ \bibinfo {pages} {024020} (\bibinfo {year} {2005})},\ \Eprint {http://arxiv.org/abs/gr-qc/0410016} {arXiv:gr-qc/0410016} \BibitemShut {NoStop}%
\bibitem [{\citenamefont {Chen}\ \emph {et~al.}(2021)\citenamefont {Chen}, \citenamefont {Deppe}, \citenamefont {Kidder},\ and\ \citenamefont {Teukolsky}}]{Chen:2021rtb}%
  \BibitemOpen
  \bibfield  {author} {\bibinfo {author} {\bibfnamefont {Y.}~\bibnamefont {Chen}}, \bibinfo {author} {\bibfnamefont {N.}~\bibnamefont {Deppe}}, \bibinfo {author} {\bibfnamefont {L.~E.}\ \bibnamefont {Kidder}}, \ and\ \bibinfo {author} {\bibfnamefont {S.~A.}\ \bibnamefont {Teukolsky}},\ }\href {\doibase 10.1103/PhysRevD.104.084046} {\bibfield  {journal} {\bibinfo  {journal} {Phys. Rev. D}\ }\textbf {\bibinfo {volume} {104}},\ \bibinfo {pages} {084046} (\bibinfo {year} {2021})},\ \Eprint {http://arxiv.org/abs/2108.02331} {arXiv:2108.02331 [gr-qc]} \BibitemShut {NoStop}%
\bibitem [{\citenamefont {Martel}\ and\ \citenamefont {Poisson}(2005)}]{Martel:2005ir}%
  \BibitemOpen
  \bibfield  {author} {\bibinfo {author} {\bibfnamefont {K.}~\bibnamefont {Martel}}\ and\ \bibinfo {author} {\bibfnamefont {E.}~\bibnamefont {Poisson}},\ }\href {\doibase 10.1103/PhysRevD.71.104003} {\bibfield  {journal} {\bibinfo  {journal} {Phys. Rev. D}\ }\textbf {\bibinfo {volume} {71}},\ \bibinfo {pages} {104003} (\bibinfo {year} {2005})},\ \Eprint {http://arxiv.org/abs/gr-qc/0502028} {arXiv:gr-qc/0502028} \BibitemShut {NoStop}%
\bibitem [{\citenamefont {Lindblom}\ \emph {et~al.}(2006)\citenamefont {Lindblom}, \citenamefont {Scheel}, \citenamefont {Kidder}, \citenamefont {Owen},\ and\ \citenamefont {Rinne}}]{Lindblom:2005qh}%
  \BibitemOpen
  \bibfield  {author} {\bibinfo {author} {\bibfnamefont {L.}~\bibnamefont {Lindblom}}, \bibinfo {author} {\bibfnamefont {M.~A.}\ \bibnamefont {Scheel}}, \bibinfo {author} {\bibfnamefont {L.~E.}\ \bibnamefont {Kidder}}, \bibinfo {author} {\bibfnamefont {R.}~\bibnamefont {Owen}}, \ and\ \bibinfo {author} {\bibfnamefont {O.}~\bibnamefont {Rinne}},\ }\href {\doibase 10.1088/0264-9381/23/16/S09} {\bibfield  {journal} {\bibinfo  {journal} {Class. Quant. Grav.}\ }\textbf {\bibinfo {volume} {23}},\ \bibinfo {pages} {S447} (\bibinfo {year} {2006})},\ \Eprint {http://arxiv.org/abs/gr-qc/0512093} {arXiv:gr-qc/0512093} \BibitemShut {NoStop}%
\bibitem [{\citenamefont {Scheel}\ \emph {et~al.}(2006)\citenamefont {Scheel}, \citenamefont {Pfeiffer}, \citenamefont {Lindblom}, \citenamefont {Kidder}, \citenamefont {Rinne},\ and\ \citenamefont {Teukolsky}}]{Scheel:2006gg}%
  \BibitemOpen
  \bibfield  {author} {\bibinfo {author} {\bibfnamefont {M.~A.}\ \bibnamefont {Scheel}}, \bibinfo {author} {\bibfnamefont {H.~P.}\ \bibnamefont {Pfeiffer}}, \bibinfo {author} {\bibfnamefont {L.}~\bibnamefont {Lindblom}}, \bibinfo {author} {\bibfnamefont {L.~E.}\ \bibnamefont {Kidder}}, \bibinfo {author} {\bibfnamefont {O.}~\bibnamefont {Rinne}}, \ and\ \bibinfo {author} {\bibfnamefont {S.~A.}\ \bibnamefont {Teukolsky}},\ }\href {\doibase 10.1103/PhysRevD.74.104006} {\bibfield  {journal} {\bibinfo  {journal} {Phys. Rev. D}\ }\textbf {\bibinfo {volume} {74}},\ \bibinfo {pages} {104006} (\bibinfo {year} {2006})},\ \Eprint {http://arxiv.org/abs/gr-qc/0607056} {arXiv:gr-qc/0607056} \BibitemShut {NoStop}%
\bibitem [{\citenamefont {Pretorius}(2005{\natexlab{b}})}]{Pretorius:2004jg}%
  \BibitemOpen
  \bibfield  {author} {\bibinfo {author} {\bibfnamefont {F.}~\bibnamefont {Pretorius}},\ }\href {\doibase 10.1088/0264-9381/22/2/014} {\bibfield  {journal} {\bibinfo  {journal} {Class. Quant. Grav.}\ }\textbf {\bibinfo {volume} {22}},\ \bibinfo {pages} {425} (\bibinfo {year} {2005}{\natexlab{b}})},\ \Eprint {http://arxiv.org/abs/gr-qc/0407110} {arXiv:gr-qc/0407110} \BibitemShut {NoStop}%
\bibitem [{\citenamefont {Okounkova}\ \emph {et~al.}(2019{\natexlab{a}})\citenamefont {Okounkova}, \citenamefont {Scheel},\ and\ \citenamefont {Teukolsky}}]{Okounkova:2018pql}%
  \BibitemOpen
  \bibfield  {author} {\bibinfo {author} {\bibfnamefont {M.}~\bibnamefont {Okounkova}}, \bibinfo {author} {\bibfnamefont {M.~A.}\ \bibnamefont {Scheel}}, \ and\ \bibinfo {author} {\bibfnamefont {S.~A.}\ \bibnamefont {Teukolsky}},\ }\href {\doibase 10.1103/PhysRevD.99.044019} {\bibfield  {journal} {\bibinfo  {journal} {Phys. Rev. D}\ }\textbf {\bibinfo {volume} {99}},\ \bibinfo {pages} {044019} (\bibinfo {year} {2019}{\natexlab{a}})},\ \Eprint {http://arxiv.org/abs/1811.10713} {arXiv:1811.10713 [gr-qc]} \BibitemShut {NoStop}%
\bibitem [{\citenamefont {Okounkova}\ \emph {et~al.}(2019{\natexlab{b}})\citenamefont {Okounkova}, \citenamefont {Stein}, \citenamefont {Scheel},\ and\ \citenamefont {Teukolsky}}]{Okounkova:2019dfo}%
  \BibitemOpen
  \bibfield  {author} {\bibinfo {author} {\bibfnamefont {M.}~\bibnamefont {Okounkova}}, \bibinfo {author} {\bibfnamefont {L.~C.}\ \bibnamefont {Stein}}, \bibinfo {author} {\bibfnamefont {M.~A.}\ \bibnamefont {Scheel}}, \ and\ \bibinfo {author} {\bibfnamefont {S.~A.}\ \bibnamefont {Teukolsky}},\ }\href {\doibase 10.1103/PhysRevD.100.104026} {\bibfield  {journal} {\bibinfo  {journal} {Phys. Rev. D}\ }\textbf {\bibinfo {volume} {100}},\ \bibinfo {pages} {104026} (\bibinfo {year} {2019}{\natexlab{b}})},\ \Eprint {http://arxiv.org/abs/1906.08789} {arXiv:1906.08789 [gr-qc]} \BibitemShut {NoStop}%
\bibitem [{\citenamefont {Cook}\ and\ \citenamefont {Whiting}(2007)}]{Cook:2007wr}%
  \BibitemOpen
  \bibfield  {author} {\bibinfo {author} {\bibfnamefont {G.~B.}\ \bibnamefont {Cook}}\ and\ \bibinfo {author} {\bibfnamefont {B.~F.}\ \bibnamefont {Whiting}},\ }\href {\doibase 10.1103/PhysRevD.76.041501} {\bibfield  {journal} {\bibinfo  {journal} {Phys. Rev. D}\ }\textbf {\bibinfo {volume} {76}},\ \bibinfo {pages} {041501} (\bibinfo {year} {2007})},\ \Eprint {http://arxiv.org/abs/0706.0199} {arXiv:0706.0199 [gr-qc]} \BibitemShut {NoStop}%
\bibitem [{\citenamefont {Owen}(2007)}]{Owen:2007dya}%
  \BibitemOpen
  \bibfield  {author} {\bibinfo {author} {\bibfnamefont {R.}~\bibnamefont {Owen}},\ }\emph {\bibinfo {title} {{Topics in numerical relativity : the periodic standing-wave approximation, the stability of constraints in free evolution, and the spin of dynamical black holes}}},\ \href@noop {} {Ph.D. thesis},\ \bibinfo  {school} {Caltech} (\bibinfo {year} {2007})\BibitemShut {NoStop}%
\bibitem [{\citenamefont {Deppe}\ \emph {et~al.}(2023)\citenamefont {Deppe}, \citenamefont {Throwe}, \citenamefont {Kidder}, \citenamefont {Vu}, \citenamefont {Nelli}, \citenamefont {Armaza}, \citenamefont {Bonilla}, \citenamefont {H\'ebert}, \citenamefont {Kim}, \citenamefont {Kumar}, \citenamefont {Lovelace}, \citenamefont {Macedo}, \citenamefont {Moxon}, \citenamefont {O'Shea}, \citenamefont {Pfeiffer}, \citenamefont {Scheel}, \citenamefont {Teukolsky}, \citenamefont {Wittek} \emph {et~al.}}]{spectrecode}%
  \BibitemOpen
  \bibfield  {author} {\bibinfo {author} {\bibfnamefont {N.}~\bibnamefont {Deppe}}, \bibinfo {author} {\bibfnamefont {W.}~\bibnamefont {Throwe}}, \bibinfo {author} {\bibfnamefont {L.~E.}\ \bibnamefont {Kidder}}, \bibinfo {author} {\bibfnamefont {N.~L.}\ \bibnamefont {Vu}}, \bibinfo {author} {\bibfnamefont {K.~C.}\ \bibnamefont {Nelli}}, \bibinfo {author} {\bibfnamefont {C.}~\bibnamefont {Armaza}}, \bibinfo {author} {\bibfnamefont {M.~S.}\ \bibnamefont {Bonilla}}, \bibinfo {author} {\bibfnamefont {F.}~\bibnamefont {H\'ebert}}, \bibinfo {author} {\bibfnamefont {Y.}~\bibnamefont {Kim}}, \bibinfo {author} {\bibfnamefont {P.}~\bibnamefont {Kumar}}, \bibinfo {author} {\bibfnamefont {G.}~\bibnamefont {Lovelace}}, \bibinfo {author} {\bibfnamefont {A.}~\bibnamefont {Macedo}}, \bibinfo {author} {\bibfnamefont {J.}~\bibnamefont {Moxon}}, \bibinfo {author} {\bibfnamefont {E.}~\bibnamefont {O'Shea}}, \bibinfo {author} {\bibfnamefont {H.~P.}\ \bibnamefont {Pfeiffer}}, \bibinfo {author} {\bibfnamefont {M.~A.}\ \bibnamefont
  {Scheel}}, \bibinfo {author} {\bibfnamefont {S.~A.}\ \bibnamefont {Teukolsky}}, \bibinfo {author} {\bibfnamefont {N.~A.}\ \bibnamefont {Wittek}},  \emph {et~al.},\ }\href {\doibase 10.5281/zenodo.10309520} {\enquote {\bibinfo {title} {\texttt{SpECTRE v2023.12.08}},}\ }\bibinfo {howpublished} {\href{https://doi.org/10.5281/zenodo.10309520}{10.5281/zenodo.10309520}} (\bibinfo {year} {2023})\BibitemShut {NoStop}%
\bibitem [{\citenamefont {Zhu}\ \emph {et~al.}(2023{\natexlab{b}})\citenamefont {Zhu} \emph {et~al.}}]{Zhu:2023fnf}%
  \BibitemOpen
  \bibfield  {author} {\bibinfo {author} {\bibfnamefont {H.}~\bibnamefont {Zhu}} \emph {et~al.},\ }\href@noop {} {\  (\bibinfo {year} {2023}{\natexlab{b}})},\ \Eprint {http://arxiv.org/abs/2312.08588} {arXiv:2312.08588 [gr-qc]} \BibitemShut {NoStop}%
\end{thebibliography}%
\clearpage
\end{document}